\documentclass[numberedappendix]{emulateapj}
\usepackage{graphicx}
\newcommand{\figwidth}{16cm}

\let\strong\textbf
\let\flag

\begin{document}

\title{3D simulations of the thermal X-ray emission from young supernova remnants including efficient particle acceleration}
\shorttitle{Thermal emission from SNRs with particle acceleration}
\shortauthors{Ferrand et al.}

\author{Gilles Ferrand\altaffilmark{1, 3}, Anne Decourchelle\altaffilmark{2} and Samar Safi-Harb\altaffilmark{1, 4}}
\altaffiltext{1}{Department of Physics \& Astronomy, University of Manitoba, Winnipeg, MB, R3T 2N2, Canada}
\altaffiltext{2} {Laboratoire AIM (CEA/Irfu, CNRS/INSU, Universit{\'e} Paris VII), CEA Saclay, b{\^a}t. 709, F-91191 Gif sur Yvette, France; anne.decourchelle@cea.fr}
\altaffiltext{3}{CITA National Fellow; gferrand@physics.umanitoba.ca}
\altaffiltext{4} {Canada Research Chair; samar@physics.umanitoba.ca}

\submitted{accepted for publication in ApJ, September 28, 2012}

\begin{abstract}

Supernova remnants (SNRs) are believed to be the major contributors to Galactic cosmic rays. The detection of non-thermal emission from SNRs demonstrates the presence of energetic particles, but direct signatures of protons and other ions remain elusive. If these particles receive a sizeable fraction of the explosion energy, the morphological and spectral evolution of the SNR must be modified. To assess this, we run 3D hydrodynamic simulations of a remnant coupled with a non-linear acceleration model. We obtain the time-dependent evolution of the shocked structure, impacted by the Rayleigh-Taylor hydrodynamic instabilities at the contact discontinuity and by the back-reaction of particles at the forward shock. We then compute the progressive temperature equilibration and non-equilibrium ionization state of the plasma, and its thermal emission in each cell. This allows us to produce the first realistic synthetic maps of the projected X-ray emission from the SNR. Plasma conditions (temperature, ionization age) can vary widely over the projected surface of the SNR, especially between the ejecta and the ambient medium owing to their different composition. This demonstrates the need for spatially-resolved spectroscopy. We find that the integrated emission is reduced with particle back-reaction, with the effect being more significant for the highest photon energies. Therefore different energy bands, corresponding to different emitting elements, probe different levels of the impact of particle acceleration. Our work provides a framework for the interpretation of SNR observations with current X-ray missions (\textit{Chandra}, \textit{XMM-Newton}, \textit{Suzaku}) and with upcoming X-ray missions (such as \textit{Astro-H}).

\end{abstract}



\section{Introduction
\label{Introduction}}

Young supernova remnants (SNRs) are a key link between stars and the Galactic interstellar medium (ISM). Understanding their morphological and spectral evolution is important to assess both explosion models of supernovae, of which SNRs are fingerprints (e.g. \citealt{Badenes2008a}) and large-scale models of the turbulent ISM, for which SNRs are seeds (e.g. \citealt{Joung2006a}).

A SNR is produced by the interaction of the material ejected during a supernova with the ambient medium \citep{Chevalier1977a}. The ejecta, being highly supersonic with typical Mach numbers initially greater than 1000,  induce a shock wave which propagates ahead of them in the surrounding medium. This shock, called the forward shock, accelerates and heats the ISM. As the ejecta push the layer of swept-up ISM, they get abruptly decelerated, inducing another shock wave that propagates backwards towards the center of the SNR. This shock, called the reverse shock, decelerates and heats the ejecta. Although the basic dynamics of the shocked structure can be studied with 1D models, only 3D simulations can properly assess various important aspects such as asymmetric explosions, mixing inside the ejecta, or interaction with a non-uniform ambient medium (in density or in magnetic field, for instance). In particular, the interface between the shocked ejecta and the shocked ISM, called the contact discontinuity, is known to be hydrodynamically unstable: because the dense ejecta are being decelerated by a medium of lower density, the interface is subject to the Rayleigh-Taylor instability, which produces a distinctive pattern of fingers and holes (see e.g. \citealt{Chevalier1992a}, \citealt{Dwarkadas2000a}, \citealt{Wang2001a}, \citealt{Blondin2001a}). It is important to know how such structures will affect images of these objects, which are only seen in projection, and how this may affect their diagnostics.

In addition to their key role in the dispersion of heavy elements synthesized inside stars and during supernovae, SNRs are widely believed to be the main production sites of Galactic cosmic-rays up to energies of about 
10$^{15}$~eV. A~fraction of the particles swept-up by the blast wave is expected to gain energy through the {\em diffusive shock acceleration} mechanism (DSA), a first-order Fermi process which relies on both the velocity discontinuity at the shock and the presence of magnetic turbulence on both sides of the shock. Whether the same mechanism could also operate at the reverse shock propagating back into the ejecta is debated \citep{Ellison2005a}. In any case, growing evidence has accumulated in the last decade making the case for SNRs being particle accelerators; see \cite{Reynolds2008a} and \cite{Vink2012a} for recent reviews on the high-energy emission from SNRs, and \cite{Ferrand2012b} for an up-to-date catalogue of X-ray and $\gamma$-ray observations of Galactic SNRs. However, a direct proof of the acceleration of hadrons in these objects is still lacking. Particles detected without ambiguity are highly energetic electrons, which radiate far more efficiently than protons. In the radio regime, the synchrotron radiation from electrons has been routinely detected for half a century, and is in fact a cornerstone for the identification of SNRs. In X-rays, the tail of this same radiation has been detected in several objects in the last 20~years, and was even found to dominate the emission in a few SNRs. In gamma-rays, several remnants have finally been detected in the last decade; the pending issue remains to disentangle the leptonic and hadronic contributions associated with Inverse Compton scattering and pion decay, respectively \citep{Gabici2008a}. It is thus important to be able to simulate the non-thermal emission from SNRs. Such work has been undertaken by \cite{Petruk2009c} and \cite{Orlando2011a}, assuming a given distribution of accelerated electrons on top of 3D MHD simulations of the remnant from SN1006. Using a model of acceleration, \cite{Lee2008a} computed the broad-band emission from both protons and electrons, on top of a 1D hydrodynamic model of the remnant's expansion.
At this point it is worth noting that, if SNRs accelerate particles as efficiently as we think, then this must affect their own dynamics in a sizeable way. If a substantial fraction of the explosion energy is diverted into the acceleration of particles at the shock, the evolution of the latter must substantially deviate from usual hydrodynamics \citep{Chevalier1983a,Decourchelle2000a}. This, in turn, should have a visible impact on the {\em thermal} emission from the remnant. Therefore, probing the morphological and spectral evolution of the SNR itself provides an indirect, but promising, way to assess the efficiency of particle acceleration in young remnants and, importantly, the acceleration of hadrons given that leptons have a negligible inertia. Indeed observational evidence for modified morphologies has been reported in the remnants of Tycho \citep{Warren2005a}, SN~1006 \citep{Cassam-Chenai2008a,Miceli2009a} and Cas~A \citep{Patnaude2009b}. Thus a few teams have begun to address the effect of back-reaction of particles on the evolution of a SNR and its impact on the thermal emission. \cite{Ellison2007a} investigated both the thermal and non-thermal emission from a spherically symmetric SNR, using a semi-analytical model for DSA \citep{Blasi2005a} and an approximate treatment for the ionization state of the plasma. Patnaude et al (\citeyear{Patnaude2009a,Patnaude2010a}) expanded on this work by making a direct coupling of an~ionization and emission code with a 1D hydrodynamical code, to study how acceleration impacts the shocked ISM behind the forward shock (the emission from the ejecta was not considered). Using this model, \cite{Castro2011a} produced broad-band spectra with both the thermal and non-thermal emission. \cite{Ferrand2010a} coupled the same kind of acceleration model to a~3D code, enabling the study of the development of the instabilities over the edge of the ejecta concomitantly with the shrinking of the shocked region caused by energetic particles. The hydrodynamic results show that the positions of the shock waves in Tycho's SNR can readily be explained by the presence of energetic ions.

Here we extend this work, by computing the emission from SNRs undergoing efficient particle acceleration. In the present paper we compute the thermal emission (the non-thermal emission will be presented in a subsequent paper). More precisely, we concentrate on the X-ray emission, which is quantitatively the largest emission coming from the hot interior of SNRs, and which is qualitatively an important source of diagnostics for these objects. Thermal X-ray emission allows for the derivation of global parameters of the remnant, such as the density and expansion velocity. Comparison of the observations with models and simulations also allows the assessment of key parameters of the acceleration process, such as injection fraction or energetics efficiency. 

The outline of the paper is as follows. In Section~\ref{sec:model}, we present our method. We recall how we perform 3D hydrodynamic simulations of the remnant evolution, coupled with a model of non-linear acceleration. We explain how these simulations are post-processed to obtain the non-equilibrium ionization state of the plasma and compute its emission. In Section~\ref{sec:results}, we present our results. We first show some useful hydrodynamic and thermodynamic quantities, then discuss synthetic spectra and projected maps from our model SNR. In Section~\ref{sec:discussion}, we summarize our main results in link with observations, and discuss the main physical assumptions underlying these. In Section~\ref{sec:conclusion}, we present our conclusions and perspectives.


\section{Model}
\label{sec:model}

In this section we present our 3D numerical model, which computes the morphological and spectral evolution of a young SNR (Section~\ref{sec:model_SNR}), including the acceleration of particles at the forward shock and the back-reaction effects (Section~\ref{sec:model_accel}).

\subsection{Morphological and spectral evolution of a young SNR}
\label{sec:model_SNR}

In our approach computations are done in two steps: first we compute the hydrodynamic evolution of the remnant (Section~\ref{sec:model_SNR_hydro}) with some assumptions on the thermodynamics of the shocked material (Section~\ref{sec:model_SNR_temp}), then we compute the ionization state of the plasma (Section~\ref{sec:model_SNR_ionization}) and its thermal emission (Section~\ref{sec:model_SNR_emission}).

\subsubsection{Hydrodynamics
\label{sec:model_SNR_hydro}}

We~consider that the SNR is initially spherically symmetric. The structure of the shocked region then depends on the radial density profiles of the ejecta ($\rho_{\rm eje}$) and of the ISM ($\rho_{\rm ism}$). We~assume power-law profiles in both media: 
\begin{equation}
\label{eq:rho_eje}
\rho_{\rm eje}(r,t) \propto r^{-n} \:t^{n-3}, 
\end{equation}
\begin{equation}
\label{eq:rho_ism}
\rho_{\rm ism}(r,t) \propto r^{-s}, 
\end{equation}
so that the radial profiles of all hydrodynamic quantities (density, velocity, pressure) are initially self-similar, and can be computed by the integration of ordinary differential equations \citep{Chevalier1982a}. Given equations~(\ref{eq:rho_eje}) and~(\ref{eq:rho_ism}), dimensional analysis shows that the radius of the contact discontinuity increases with time as
\begin{equation}
\label{eq:r_CD}
a(t) = t^{(n-3)/(n-s)}, 
\end{equation}

These solutions are valid until the reverse shock reaches the central core of the ejecta. After the initial ejecta-dominated phase, the SNR enters another phase, still adiabatic but now dominated by the dynamics of the swept-up ISM, which is commonly called the Sedov phase. To track the evolution of the SNR over time, without any limiting assumptions, we make time-dependent numerical simulations, where the profiles are initialized with the self-similar solutions and then evolve freely. We use the code RAMSES, which contains a Eulerian hydrodynamical solver based on a second-order Godunov method \citep{Teyssier2002a}. We run it with the MUSCL scheme and an exact Riemann solver. The physical domain is discretized on a Cartesian grid. As we want to catch the development of the Rayleigh-Taylor instabilities, we use a 3D grid. As the SNR can expand by several orders of magnitudes, and most interesting features are present inside the shocked region, we use a comoving grid, in which the system of waves is quasi-stationary. To obtain this, we perform a change of variables based on appropriate powers of the scale factor defined by equation~(\ref{eq:r_CD}), and we add a non-inertial force to the Euler equations (for more details see \citealt{Fraschetti2010a}). The resolution of the grid around the shocks is further increased by using RAMSES tree-based adaptive mesh refinement (AMR).

\subsubsection{Thermodynamics
\label{sec:model_SNR_temp}}

The shocked plasma (both the ISM downstream of the forward shock, and the ejecta downstream of the reverse shock) is heated to such high temperatures that it shines in X-rays. In~the reference case without any losses, that is, as long as a negligible amount of energy is channeled into energetic particles or lost through radiation,\footnote{Note that although in this paper we are computing the radiation from the remnant, we restrict our study to the so-called non-radiative phases of its evolution, meaning that radiation has no impact yet on its dynamics. For cosmic-rays the situation is reversed: they are thought to be mostly produced during these early phases, and thus are fully accounted for in our approach.} the average temperature immediately downstream of the shock ($T_{\rm S}$)  depends solely on the shock speed ($v_{\rm S}$) according to:
\begin{equation}
\label{eq:T2_vS}
T_{\rm S} = \frac{3}{16} \frac{\bar{m}}{k_B} v_{\rm S}^2 \sim 2\times10^7\;{\rm K}\;\left( \frac{v_{\rm S}}{1000\;{\rm km/s}} \right)^2
\end{equation}
where $\bar{m}=\mu m_p$ is the mean molecular weight ($m_p$ being the proton mass and $\mu$ the weight factor accounting for the composition of the plasma) and $k_B$ is the Boltzmann constant. 

The emissivity of the plasma  depends on the temperature of the electrons ($T_e$) which excite and ionize ions. It~is useful to introduce the ratio~$\beta$ of the electrons and protons temperatures:
\begin{equation}
\label{eq:beta}
\beta = \frac{T_e}{T_p},
\end{equation}
which is a function of time and position. The value of this ratio at the shock front ($\beta_{\rm S}$) is poorly unknown, and is therefore a free parameter of the model. Theoretically, it could be anywhere between the masses ratio ($\beta_{\rm S} = m_e/m_p$) and full equilibration ($\beta_{\rm S} = 1$). Observations suggest prompt equilibration for shock speeds below about 400~km/s, but intermediate degrees of equilibration for faster shocks, with $\beta_{\rm S}$ tending towards zero roughly as $v_{\rm S}^{-2}$ (see \citealt{Rakowski2005a} and \citealt{Ghavamian2007a} for reviews).
Downstream of the shock, electrons will be heated through Coulombian interactions with ions, so that $T_e$ will progressively equilibrate with $T_p$.\footnote{Assuming that ions species are heated according to their mass, and given that their various masses are not widely different (compared to the electron mass), we consider that all ion species quickly get roughly the same temperature after crossing the shock front, and focus only on the progressive equilibration of electrons with ions.} To compute $T_e$ in any given cell in the shocked region, we use the method of \cite{Cox1982a}, based on the model of \cite{Itoh1978a,Itoh1979a}. This model assumes an adiabatic evolution of the plasma. It provides the current value of the temperature ratio $\beta$, given its value at the shock~$\beta_{\rm S}$, the current density~$n$ and temperature~$T$ of the cell, and its age~$\tau_{\rm S}$ defined as
\begin{equation}
\label{eq:tau_S}
\tau_{\rm S}\left(\mathbf{x},t\right) = \int_{t_{\rm S}}^t {\rm d}t'
\end{equation}
where $t_{\rm S}$ is the time when the material considered was shocked. In the numerical simulations, we keep track of $\tau_{\rm S}$ with the inclusion of a passive scalar in RAMSES, which is incremented in the shocked region only.

\subsubsection{Non-equilibrium ionization
\label{sec:model_SNR_ionization}}

Free electrons in the plasma excite and ionize the atoms, until an equilibrium is reached between ionization and recombination for all species. We use ionization and recombination rates from \cite{Arnaud1985a}, \cite{Arnaud1992a}, and \cite{Mazzotta1998a}. The final ionization state depends only on the electron temperature~($T_e$), but remnants we consider here are much too young to have reached ionization equilibrium, so that the ionization state also depends on the electron density ($n_e$) and on the time ($t$) elapsed since the plasma has been shocked. Ionization timescales for various elements, as a function of temperature, are given in \cite{Smith2010a}. It~is apparent that no species heavier than Carbon can be ionized for $n_e \times t < 10^{7}~{\rm cm^{-3}.s}$, and that equilibrium of all ions is granted for $n_e \times t > 10^{13}~{\rm cm^{-3}.s}$. In between, the actual ionization state must be computed. 
The value of the electron density, $n_e$, at any time, $t$, depends on the ionization state of the plasma, so that the computation of its ionization state should in principle be coupled with its hydrodynamic evolution at each time step. We simplify the problem by deferring the computation of the actual~$n_e$ to the end of the simulation, recording only the evolution of the total density over time. The hydrodynamic code natively works with the mass density, $\rho$, so we added another passive scalar, $\tau_{\rm I}$, in RAMSES defined as: 
\begin{equation}
\label{eq:tau_I}
\tau_{\rm I}\left(\mathbf{x},t\right) = \int_{t_{\rm S}}^t \rho \left(\mathbf{x},t'\right).{\rm d}t'
\end{equation}
where we recall that $t_{\rm S}$ is the time when the material considered was shocked (note that we use the convention of denoting quantities defined since the shock crossing with Greek letters). So after material is shocked, we keep track of its ionization age as it is advected downstream, and at the end of the simulation we compute the non-equilibrium ionization state in each cell. To do so we use the so-called exponentiation method, introduced independently by \cite{Masai1984a} and \cite{Hughes1985a}, and reviewed in \cite{Smith2010a}. In~general, the ionization evolution of an element of atomic number~$Z$ is given by a coupled set of $Z+1$ equations; in~this method, these are recast as $Z$ uncoupled equations, after diagonalization of the rates matrix. These $Z$ first-order differential equations are integrated to obtain the ionization fractions of the element, for a given electronic temperature ($T_{\rm e}$) and ionization age ($\int{n_{\rm e}{\rm d}t}$). The ionization age is defined implicitly, as $n_{\rm e}$ depends on the ionization fractions, which can only be computed for some value of $\int{n_{\rm e}{\rm d}t}$, hence the need for an iterative computation. A first guess for the ionization age is obtained from the tracer $\tau_{\rm I}$, assuming that all species are ionized once -- an assumption that is sufficient for the shocked ISM, where Hydrogen dominates, but not for the shocked ejecta, where heavy elements are not negligible. This value is used to compute the ionization fractions, from which the actual value of $n_{\rm e}$ is obtained. Then the ionization fractions are re-computed, and the thermal emission is computed. 

\subsubsection{X-ray thermal emission
\label{sec:model_SNR_emission}}

The hot ionized plasma located between the two shocks of a young SNR is a source of photons with energies of the order of keV, which is now routinely observed with space-based X-ray instruments (see e.g. \citealt{Decourchelle2005a} for a review). This X-ray emission is a function of the electronic density and temperature of the plasma, and of the ionization fractions of each species (which themselves depend on density and temperature as explained in the previous section). We~compute this thermal emission as a function of the photon energy with a spectral code from \cite{Mewe1985a,Mewe1986a}. The code computes both the continuum emission (free-free, free-bound and two-photons radiations) and the emission lines (from excitation and recombination processes), for each of the following 15~elements: Hydrogen (H, $Z=1$), Helium (He, $Z=2$), Carbon (C, $Z=6$), Nitrogen (N, $Z=7$), Oxygen (O, $Z=8$), Neon (Ne, $Z=10$), Sodium (Na, $Z=11$), Magnesium (Mg, $Z=12$), Aluminium (Al, $Z=13$), Silicon (Si, $Z=14$), Sulfur (S, $Z=16$), Argon (Ar, $Z=18$), Calcium (Ca, $Z=20$), Iron (Fe, $Z=26$), and Nickel (Ni, $Z=28$).

The total thermal emission from any given cell is the sum of the emission from the ejecta and from the ISM. The distribution of the ejecta is known thanks to another scalar variable passively advected with the fluid density, the fraction of ejecta~$f$, defined so that the ejecta density is $f \times \rho$ and the ISM density is $(1-f) \times \rho$. The thermal emission of species~$i$ is proportional to $n_e \times n_i$, with $n_e$ the electrons density and $n_i$ the ion density, which are computed from~$\rho$ according to the abundances and ionic fractions. The ejecta and the ISM are distinguished by their composition: we assume solar abundances for the ISM (we use standard values from \citealt{Anders1989a}), and metal-enhanced abundances for the ejecta (we adopt the model of \citealt{Maeda2010b}).\footnote{Other composition models are available for the ejecta (see for instance the reference work by \citealt{Nomoto1984a}, and a more recent work by \citealt{Badenes2005b}), but comparing them is beyond the scope of this paper.} For reference, this corresponds to molecular weights of $\mu\simeq1.9$ in the ejecta and $\mu\simeq0.6$ in the ISM, if we assumed full ionization (the most abundant, light species in the ISM are indeed likely to be fully ionized; the most abundant, heavy species in the ejecta are not likely to be).
Note that, in each medium, we do not consider spatial variations in the abundances, as in this first study of the thermal emission from 3D simulations, we are mostly interested in disentangling projection effects, independently of the supernova which gave birth to the structure. 

The emission is computed in each shell of the shocked region between the reverse and the forward shocks. It is then projected along one axis of the cartesian grid, properly taking into account the tree-based AMR structure. We thus obtain 2D synthetic maps of the emission of the remnant. Note that, even though our final product is 2D, it is mandatory to make 3D simulations to properly capture the development of the Rayleigh-Taylor instabilities at the contact discontinuity. This is all the more important since the thermal emission is dominated by the shocked ejecta which exhibit a complicated morphology.

\subsection{Acceleration and back-reaction of particles}
\label{sec:model_accel}

Strong and magnetized shocks, such as the ones encountered in young SNRs, can accelerate particle to high energies. It is not the purpose of this paper to study the fate of these particles in detail, and in particular their \emph{non-thermal} emission which will be presented in a subsequent paper. Our point here is to show how the \emph{thermal} emission from the SNR is affected by particle acceleration. In this section we explain how we model acceleration at the shock (Section~\ref{sec:model_accel_DSA}), how we handle the back-reaction of accelerated particles on the flow (Section~\ref{sec:model_accel_back}), and how this affects the diagnostics needed to compute the thermal emission (Section~\ref{sec:model_accel_diag}). Note that, although our model is general, we choose to use it at the forward shock only, because it is unclear whether the reverse shock can be also an efficient accelerator (see more in the discussion, Section~\ref{sec:discussion-reverse-shock}).

\subsubsection{Diffusive shock acceleration
\label{sec:model_accel_DSA}}

A~fraction of the particles swept-up by the blast wave is expected to be accelerated through the diffusive shock acceleration mechanism \citep{Drury1983a,Malkov2001c}.
In~the test-particle regime, this mechanism produces a power-law energy spectrum with an index ($s$) that is universal, because it is entirely determined by the compression ratio ($r$) of the shock, which for strong shocks in the ISM is always $r=4$. However, as the pressure of particles builds up upstream of the shock, the gas is pre-accelerated, hence the formation of a \emph{precursor}. As the distance a particle can travel ahead of the shock is a function of its energy, particles of different energies sample different parts of the precursor, where they experience different velocity jumps, and therefore the slope~$s$ becomes energy-dependent: spectra get concave. Solving explicitly the coupled system of the fluid and the particles as a function of time can be done numerically (e.g. \citealt{Ferrand2008a,Zirakashvili2011a}), but the high numerical cost limits the possibility to run 3D models of realistic SNRs. Instead, we use the semi-analytical model of acceleration of Blasi (\citeyear{Blasi2002a}; \citeyear{Blasi2004a}). Given the shock properties (velocity and upstream conditions), this model jointly solves the particle spectrum and the fluid velocity profile as functions of the momentum of particles. Note that it does not solve the time evolution of the system, but only finds quasi-stationary solutions, so that we have to rerun it after each hydro time-step. This procedure is justified at most energies, but breaks down close to the highest energies when the acceleration time gets of the same order as the age of the SNR (the acceleration time is computed with the classical formula derived by \cite{Drury1983a} in the linear regime, but using the modified velocities upstream and downstream of the shock to account for non-linear effects). The maximum energy can also be limited by the lack of geometrical confinement: the model includes self-consistently the escape of particles upstream of the shock, which carry energy away from the system \citep{Blasi2005a}. Assuming a uniform explosion in a uniform ISM, we expect the shock properties to be symmetric, and so at any given time we run the acceleration model only once, using azimuthally averaged shock properties. The shock fronts are identified in the hydro simulation through their strong pressure gradients.

Even though the magnetic field does not play a dynamical role in the evolution of the SNR (and is thus not activated inside RAMSES), the nature and the level of magnetic turbulence in the vicinity of the shock is crucial to particle acceleration. Both the age-limited and size-limited maximum energies of particles are computed for a given diffusion coefficient $D$. We make the common assumption of Bohm diffusion, that is
\begin{equation}
\label{D_Bohm}
D(p) = D_0 \frac{p^2}{\sqrt{1+p^2}} \quad {\rm with} \quad D_0 = \frac{3.10^{22}~{\rm cm^2.s^{-1}}}{B(\mu{\rm G})}
\end{equation}
where $p$ is the momentum of the particles in $m_{\rm p}c$ units and~$B$ is the intensity of the magnetic field. Both theoretical and observational evidence suggest that energetic particles can substantially amplify $B$ upstream of the shock (see \citealt{Bykov2012a} for a review), but we do not include this effect in our model at this point because of the large uncertainties and complexities in the mechanisms at work; we will address this issue in a subsequent paper where we compute the non-thermal emission of accelerated particles (see \cite{Lee2012a} for other recent numerical developments on this issue). However, we here include turbulent heating in the precursor \citep{Amato2006a}: we assume that Alfv{\'e}n waves are efficiently damped in the precursor, so that the gas gets compressed non-adiabatically. This tempers the intensity of the back-reaction, so that the particle pressure does not get unreasonably high.\footnote{If, instead, we had assumed that amplified Alfv{\'e}n waves are advected to the shock, the jump relations would be modified, providing another kind of regulation of the shock structure \citep{Caprioli2008a}. The degree of modification of the SNR would be of the same order, although a little higher. As far as the thermal emission is concerned, choosing a different recipe for magnetic field amplification would amount to choosing a slightly different value for the injection fraction (for the effect of $\eta$ on the shock modification, see Fig.~2 of \citealt{Ferrand2010a}). Implications for the non-thermal emission will be discussed in a subsequent paper.}

\subsubsection{Particle back-reaction
\label{sec:model_accel_back}}

When it comes to the SNR evolution, what matters foremost is the modification of the shock compression. As a precursor is formed,\footnote{In our hydrodynamic simulations we do not explicitly handle this precursor upstream of the shock. Its impact on the X-ray emission is likely very small. It can however be detected in the optical domain as if affects the Balmer lines of Hydrogen and the forbidden lines of heavier elements (e.g.~\citealt{Rakowski2011a}).} the shock discontinuity is reduced to what is called the sub-shock, of compression ratio $r_{\rm sub}<4$, whereas the overall compression (from far upstream to downstream) is increased, with a compression ratio $r_{\rm tot}>4$. To take that effect into account in our simulations, we use a pseudo-fluid with variable adiabatic index~$\gamma$,  as we already did in~\cite{Ferrand2010a} (we recall that $\gamma$ is defined so that
\begin{equation}
\label{eq:polytrope}
P \propto \rho^\gamma,
\end{equation}
\begin{equation}
\label{eq:e_int}
\epsilon=\frac{P}{\gamma-1},
\end{equation}
where $\rho$, $P$, $\epsilon$ are the fluid's mass density, pressure and internal energy, respectively).
In our version of RAMSES, $\gamma$ is a variable of each cell, and is modified in order to account for the presence of energetic particles. The pseudo-fluid thus represents both the thermal (shocked fluid) and non-thermal (accelerated particles) populations. To enforce the back-reaction of accelerated particles on the SNR structure, we thus proceed as follows: at each time-step, we compute the index ($\gamma_{\mathrm{eff}}$) which will produce the same ratio ($R_{\mathrm{tot}}$) as predicted by the acceleration model, and affect it to the cells located just upstream of the shock front. $\gamma_{\mathrm{eff}}$~thus drops down from the fiducial value of~5/3, close to the value of~4/3 appropriate for relativistic particles, or even slightly below that due to the escape of particles.\footnote{A more rigorous treatment would be to split the two effects: lower $\gamma_{\mathrm{eff}}$ to account for the presence of relativistic particles, add an energy sink to account for the escape of some of these particles. This was done by \cite{Patnaude2009a}, who observed that such changes are important for the later stages of the SNR evolution, but not at the age considered here. A~significant improvement over our current model would be a multi-fluid approach, where each population (thermal and non-thermal) has its own~$\gamma$.} Then $\gamma_{\mathrm{eff}}$ is advected downstream of the shock, inside the shocked region, so that each fluid element remembers modifications induced by particle acceleration at the time it was shocked. This modification of $\gamma$ makes the fluid more compressible, so that the shocked region gets denser and narrower \citep{Decourchelle2000a}.

The importance of back-reaction effects obviously depends on the amount of energy channeled into energetic particles, or alternatively on the fraction of particles that enter the acceleration process at the shock front. The latter mechanism, known as the \emph{injection}, is unfortunately poorly known, and therefore often parametrized.\footnote{It is however not the case in Monte-Carlo simulations of the kind of \cite{Ellison1984a}.} We adopt here the thermal leakage recipe of \citep{Blasi2005a} and set the fraction~$\eta$ of particles crossing the shock entering the acceleration process to be 
\begin{equation}
\label{eq:eta_xi}
\eta = \frac{4}{3\pi^{1/2}}\left(r_{\rm sub}-1\right)\xi^3\exp(-\xi^2)
\end{equation}
where $r_{\rm sub}$ is the compression ratio at the (sub-)shock (see next section) and $\xi$ is a free parameter parametrizing the energy of swept-up particles, defined as 
\begin{equation}
\label{eq:p_inj}
\xi = \frac{p_{\rm inj}}{p_{\rm th}}
\end{equation}
where $p_{\rm inj}$ is the injection momentum of particles and $p_{\rm th}$ is the mean thermal momentum immediately downstream of the shock. $\xi$ is expected to be in the range 2--4. The fraction~$\eta$ obtained from Eq.~\ref{eq:eta_xi} is applied to the total mass flux crossing the shock, regardless of its detailed composition: the acceleration model only asks for the proton number density, which is a reasonable simplification for the ISM swept-up by the forward shock.\footnote{See \cite{Caprioli2011a} for recent developments of the model regarding the impact of heavy species on the diffusive acceleration process.}

Finally, note that back-reaction effects might be already significant at the small age at which we initialize the simulation using self-similar profiles. The pressure of relativistic particles, computed from the same acceleration model, is therefore included in the computation of these profiles \citep{Chevalier1983a}. 

\subsubsection{Properties of the modified plasma
\label{sec:model_accel_diag}}

As already explained, the emission at each point depends on the electrons density ($n_e$) and temperature ($T_e$). These quantities are derived from the mass density ($\rho$) and the protons temperature ($T_p$) of the hydro code as explained before. But because of our use of a pseudo-fluid to implement the back-reaction of particles, we need to extract the properties of the \emph{thermal} component only.
The density of the pseudo-fluid is within a very good approximation the same as the density of the thermal fluid, given the low fraction of particles accelerated at the shock. 
Regarding temperature, a distinction must be made between the ejecta and the ISM. In the shocked ejecta, the pressure of the pseudo-fluid is actually the pressure of the real fluid, given that we do not consider acceleration at the reverse shock. Hence the temperature in the shocked ejecta~($T_{\rm ej}$) can simply be obtained as:
\begin{equation}
\label{eq:T_ej}
T_{\rm ej} = \frac{\mu m_p\;P}{\rho\;k_B}.
\end{equation}
In the shocked ISM, the pressure of the pseudo-fluid contains an important contribution from energetic particles -- in fact, it can easily be dominated by it -- so that we cannot use the same method. We determine the temperature in the shocked ISM~($T_{\rm ISM}$) a posteriori, using adiabatic evolution downstream of the shock:
\begin{equation}
\label{eq:T_ISM}
T_{\rm ISM} = T_{\rm S} \times \left(\frac{n}{n_{\rm S}}\right)^{\gamma-1}
\end{equation}
where $T=T\left(r,t\right)$ and $n=n\left(r,t\right)$ are the temperature and density of a cell at radius~$r$ and time~$t$, and $T_{\rm S}=T\left(r_{\rm S},t_{\rm S}\right)$ and $n_{\rm S}=n\left(r_{\rm S},t_{\rm S}\right)$ are the temperature and density immediately downstream of the forward shock at radius~$r_{\rm S}$ at the time it was shocked~$t_{\rm S}$. We recall that each cell knows how much time~$\tau_{\rm S}$ elapsed since its material was shocked (Eq.~\ref{eq:tau_S}). We~also need to keep track of the history of shock modifications: the values of the density~$n_{\rm S}$ and temperature~$T_{\rm S}$ just downstream of the shock, obtained from the acceleration model, are recorded at each time step.

Knowing the hydrodynamic and thermodynamic properties of the thermal plasma, we can compute its ionization state and thermal emission as explained before, now taking into account the back-reaction of accelerated particles on the SNR.

\subsection{Parameters}
\label{sec:model_parameters}

Our goal in this paper is not to make a detailed modeling of a specific object, but rather to point out the important physical aspects to consider when interpreting observations. So we do not to explore the full parameter space, but rather use a single set of representative parameters (as in our preliminary work shown in \citealt{Ferrand2010a}). 

We are considering the remnant of a type~Ia event, and are using Tycho's SNR as a reference, but we are not attempting a~fit of this object. We set the explosion energy to $E=10^{51}~{\rm erg}$, and the ejecta mass to $M_{\rm ej}=1.4$~solar masses, and adopt a power-law profile with index $n=7$. We assume a uniform ambient medium (that is, a power-law profile with index $s=0$) of rather low density $n_{H,0}=0.1\,{\rm cm^{-3}}$. We use these profiles to initialize the hydrodynamic code at a small age of $t=10$~years. We then let the code evolve the profiles until $t=500$~years. Particles are accelerated continuously in time at the forward shock, for the injection level we take the common choice $\xi=3.5$, which typically produces injection fractions of the order of a few~$10^{-4}$ for the remnant considered. The injection level is expected to vary with the obliquity angle between the mean magnetic field and the shock normal, but whether parallel or perpendicular shocks are more efficient accelerators is still debated, and the geometry of the magnetic field around the blast wave is not well known anyway. So we do not consider here the obliquity dependence of the injection, and instead use this fraction as an average value for the whole shock surface, at any given time. This assumption of overall symmetry seems reasonable for an object like Tycho (for a clearly bipolar object like SN 1006 we refer the reader to the work by \citealt{Orlando2007a} and \citealt{Petruk2009a,Petruk2009b}). As time increases, the injection fraction is lowered as $r_{\rm sub}$ is reduced, see Eq.~\ref{eq:eta_xi}. We assume an ambient magnetic field $B_0 = 5~\mu{\rm G}$ for the diffusion, leading to maximum energies of a few~$10^5\:m_p c^2$ for protons. 

Hydrodynamic results are processed at the final age to compute the thermodynamic and ionization state of the plasma in each cell. The only free parameter introduced at this stage is the ratio~$\beta_{\rm S}$ of electron and proton temperatures at the shock front, which for the sake of simplicity we fix here to~$\beta_{\rm S} = m_e/m_p \simeq 5.5\times10^{-4}$ (in the case of Tycho, observations reported by \citealt{Ghavamian2001a} suggest $\beta_{\rm S} \lesssim 20\%$ at the forward shock). 

We then compute the thermal emission in each cell. There is no parameter to tune in the emission code. We just have to define the energy grid of photons, which we choose to be logarithmic, with 100~points per decade over 2~decades in energy, from 0.1~keV to~10~keV -- which is wide enough to contain most of the X-ray emission observed by current instruments. We recall that the emission is computed only in the shocked region, as bounded by the (average) positions of the reverse and forward shock, and then projected along the line of sight. Spatially, we use an AMR grid with a formal resolution of $1024^3$, so that we produce emission maps of $1024^2$ pixels in each energy bin. The effect of numerical resolution is shown in Appendix~\ref{sec:app_resolution}. Note that we only simulate one octant of the remnant, and therefore obtain projected maps of one quarter of the remnant.


\section{Results}
\label{sec:results}

In this section we present a variety of synthetic emission maps and spectra generated from our 3D simulations, in order to investigate the effects of acceleration on observational diagnostics of the SNR. In all the plots, we compare data obtained without including the back-reaction of accelerated particles (cases labeled `OFF'), that is when forcing the (test-particle) linear regime, so that the shock structure is un-modified; and data obtained including the back-reaction of accelerated particles (cases labeled `ON'), that is when letting the non-linear regime develop, so that the shock structure is modified. 
We first show the quantities on which the thermal emission relies (Section~\ref{sec:results_diag}), before showing the emission itself (Section~\ref{sec:results_emission}).

\subsection{Plasma state}
\label{sec:results_diag}

Here we show the quantities, diagnosed from the RAMSES code, needed to compute the emission: hydrodynamic quantities (Section~\ref{sec:results_diag_hydro}), thermodynamic quantities (Section~\ref{sec:results_diag_thermo}), and ionization state (Section~\ref{sec:results_diag_ionis}). We present all these data in the form of slices (in a plane $z=0$ perpendicular to the line of sight) in order to show the inner structure of the remnant.

\subsubsection{Hydrodynamics
\label{sec:results_diag_hydro}}

\begin{figure}
\renewcommand{\figwidth}{16cm}
\centering
\includegraphics[width=\figwidth]{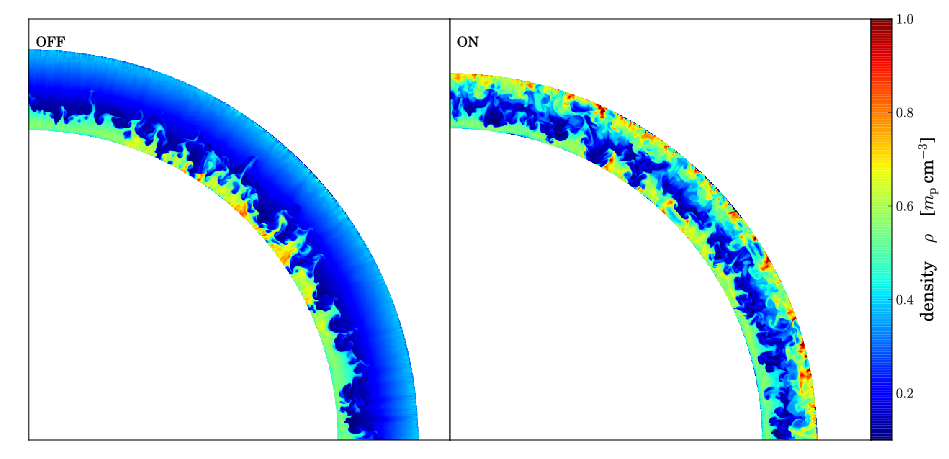}
\includegraphics[width=\figwidth]{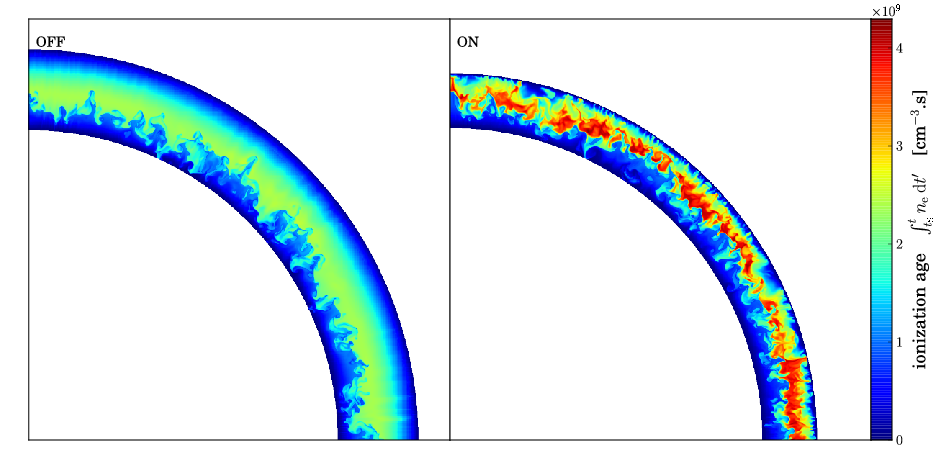}
\caption{Slices of hydrodynamic quantities, in the $z=0$ plane, in the shocked region.
Top: mass density, 
bottom: ionization age (time since the material was shocked, weighted by the electron density).
Cases without (`OFF', on the left) and with (`ON', on the right) back-reaction of particles are compared.
\label{fig:cut_diag_hydro}}
\end{figure}

The mass density~$\rho$ is shown at the top panel of Figure~\ref{fig:cut_diag_hydro}. It~ranges over one order of magnitude, from roughly $0.1$ to $1$~$m_p$.cm$^{-3}$. The fluid consists of different species, the density shown here is the sum of all nucleons over all elements. The fluid also consists of two media, the ejecta in the inner region and the ISM in the outer region. In between the two we observe the development of the Rayleigh-Taylor instabilities: fingers of ejecta are protruding inside the ISM, with mushroom-shaped heads that sometimes get disrupted. The contact discontinuity between the ejecta and the ISM gets a very irregular shape, and is often ill-defined as the two media get mixed. The bottom panel of the same Figure~\ref{fig:cut_diag_hydro} shows the ionization age~$\int{n_e.{\rm d}t}$ of the material (see Section~\ref{sec:model_SNR_ionization}). In the ISM, this age mostly reflects the time since the material was shocked, rising progressively from the forward shock to the contact discontinuity, which is highlighted by a sudden drop in density. The interface is much distorted though, so that old material can be found in a wide range of radii within the shocked region -- a feature which makes 3D simulations mandatory. In the ejecta, the ionization age remains lower, even though they are denser, because heavier elements take more time to get ionized after they get shocked by the reverse shock.
With our parameters, the maximum ionization age is $\simeq 2.5 \times 10^9$~cm$^{-3}$.s in the un-modified case. According to the discussion of Section~\ref{sec:model_SNR_ionization}, it is apparent that the plasma must be in a non-equilibrium state.

\flag{The effect of acceleration} can be seen when comparing the left and right sides of the figure: as part of the energy of the system is diverted into the acceleration of particles, the shocked region shrinks. This means two things. First, it gets narrower in radius, here by about one third, which already makes an observational diagnostics \citep{Ferrand2010a}. Second, it gets denser close to the forward shock (where acceleration happens), here by  a factor of about~2. Whereas in the un-modified case the highest densities are found in the ejecta, close to the reverse shock, in the modified case the same kind of values are obtained next to the reverse and forward shocks. The region immediately downstream of the forward shock looks irregular when back-reaction is activated, but we warn the reader that at least part of it is caused by numerical instabilities triggered at the shock front. These instabilities, commonly known as {\em carbuncle}, develop because of the quasi-stationarity of the shock in the comoving grid \citep{Fraschetti2010a}, and are exacerbated when the shock is modified by particles.\footnote{Although unphysical features could be smoothed with various numerical strategies (including using more diffusive Riemann solvers, or using a wobbling comoving grid), it appears difficult to completely avoid them while maintaining a sufficient resolution of the physical structures inside the shocked region. As a result, our current numerical setup does not allow us to discuss the azimuthal profile of the shock.} The overall increase in density close to the shock, however, is perfectly real. And part of these post-shock turbulent features might in fact be real, and related to Rayleigh-Taylor structures, as the radial density profile is substantially modified in the shocked ISM. At any rate, the exact density profile close to the forward shock should not much affect the maps of thermal emission presented later, as long as the ejecta dominate, which is usually the case, and even more so when including back-reaction, so that we can pursue our study. As we do not include acceleration at the reverse shock, the region between the reverse shock and the contact discontinuity is not much affected by acceleration; however, it seems that the ejecta cannot protrude as far away in the ISM, as they approach the closer and denser outer shock edge. Because the shocked ISM is more compressed, the ionization age gets higher here, by almost a factor of~2. 

\subsubsection{Thermodynamics
\label{sec:results_diag_thermo}}

Temperatures are shown on Figure~\ref{fig:cut_diag_T} for protons and for electrons (see Section~\ref{sec:model_SNR_temp}). As expected, the temperature is higher in the center of the shocked region, close to the contact discontinuity, where material was shocked at an earlier age (compare with bottom of Figure~\ref{fig:cut_diag_hydro}), when the shock speed was much higher. Over the course of the simulation, the speed of the blast wave in the ISM falls down from $\simeq 30,000$ km.s$^{-1}$ at the initial time $t=10$~yr to $\simeq 5000$~km.s$^{-1}$ at the final time $t=500$~yr. In the un-modified case, the protons temperature is of the order of $10^9$~K (see Eq.~\ref{eq:T2_vS}). The electrons temperature (the one that matters regarding thermal X-ray emission) is much lower (by more than two orders of magnitude), given our assumptions that heating at the shock depends on the mass ratio and equilibration with protons is progressive. $T_{\rm e}$~can still reach about $20\times10^6$~K, which is enough to produce intense X-ray emission. 

\flag{The effect of acceleration} is important for the ISM, given our assumption of efficient acceleration at the forward shock. Because a significant part of the energy goes into acceleration of particles, the temperature of protons $T_{\rm p}$ is much lower in the shocked ISM with back-reaction: by typically a factor of~4 in the case simulated here. This effect is well known, see e.g.~\cite{Decourchelle2000a} for an early 1D modeling and \cite{Hughes2000a} for a possible observational diagnostic. $T_{\rm p}$~is also indirectly reduced in the shocked ejecta, although by usually less than 10\%. The temperature of electrons is not as strongly affected by particle acceleration: with back-reaction, $T_{\rm e}$ is lowered by about a factor of~2 in the shocked ISM (it remains the same within about 1\% in the shocked ejecta). The different behaviour of electrons and protons with respect to particle acceleration stems from our hypothesis of progressive equilibration downstream of the shock (see Section~\ref{sec:model_SNR_temp}). Even though the ratio $\beta = T_{\rm e}/T_{\rm p}$ ends up being much higher than the initial value $m_e/m_p$ assumed at the shock, it is far from reaching unity, no matter how low $T_{\rm p}$ gets: in the ISM it reaches about 1\% in the un-modified case and about 2\% in the modified case. 

\subsubsection{Ionization
\label{sec:results_diag_ionis}}

Knowledge of the density, density-weighted age, and electron temperature of the plasma allows us to compute the ionization state of the plasma a posteriori, as explained in Section~\ref{sec:model_SNR_ionization}. On Figure~\ref{fig:cut_diag_ne} we show the total electron density. Comparing with the top of Figure~\ref{fig:cut_diag_hydro}, it can be seen that in the ISM, which mostly consists of light, fully ionized, species, there is close to one free electron per nucleon; whereas in the ejecta, there are many less free electrons, as heavy species take time to lose their electrons. We will not discuss in detail the precise morphology of all ionization states of all species, but instead present an average view of the state of a few important ones. On Figure~\ref{fig:cut_diag_Z} we show maps of the mean electric charge $<Z>$ of Oxygen, Silicon, and Iron. $<Z>$ is defined as the sum of the charges of all the ionization states of a given element, weighted by their respective density fractions. It starts at $<Z>=1$ at the shock fronts (by assumption), and quickly rises behind the shocks to reach a kind of plateau, at a value dependent on the total charge of the atom. The more electrons an element has, the less chances it has to approach full ionization. Superimposed on this global radial evolution, Rayleigh-Taylor structures are clearly visible. 

\flag{The effect of acceleration} on the mean electric charge mostly stems from the higher compression of the shocked ISM: with back-reaction, the region of highly ionized states starts closer to the forward shock, which is closer to the contact discontinuity. It is worth recalling that we only present here an average quantity $<Z>$, which can mask spatial variations in the fraction of different levels of ionization of a given element. Local variations in the plasma state, and the impact of acceleration, will become more apparent when looking at the thermal emission pixel by pixel.

\subsection{Thermal X-ray emission}
\label{sec:results_emission}

Using the data shown in the previous section, the emission code computes the thermal emission in each cell, which is then projected along the line of sight (the z-axis) to generate synthetic maps. This allows us, \textit{for the first time}, to present realistic maps of the thermal emission from a SNR, including the effects of hydrodynamic turbulence and of particle acceleration. We first show global maps and spectra over the broad energy band simulated (Section~\ref{sec:results_emission_broad}); then we show the emission maps (and equivalent width maps when applicable) for several narrow energy bands (Section~\ref{sec:results_emission_narrow}), chosen to illustrate the behaviour of some prominent emission lines and of the high-energy continua. Such maps can be compared with observations made with modern and future X-ray instruments allowing for spatially resolved spectroscopy.

\subsubsection{Broad band emission}
\label{sec:results_emission_broad}

\strong{Maps}. The broad-band emission from 300~eV to 10~keV is shown on Figure~\ref{fig:prj_THtt_broad} (the emission is computed in the code from $100$~eV, but below $300$~eV interstellar absorption will prevent instruments to observe the theoretical fluxes and features simulated). The~emission includes the emission of all species (from Hydrogen to Nickel) and in all media (both the ejecta and the ISM). 
At the top of the figure, we show the emission in slices (as in the previous sections), to show where it actually comes from inside the shocked region: mostly from the Rayleigh-Taylor fingers of ejecta, plus a small and rather uniform contribution of the ISM. At the bottom of the figure, we show projected maps (the novelty of this section) obtained by summing all the slices along the line of sight. A~distinctive `fleecy' structure is obtained in projection within the remnant, because of the Rayleigh-Taylor instabilities. Again, this demonstrates the need for 3D simulations to obtain realistic maps. We note that the outer edge of the ejecta is always brighter than the interior of the remnant, because more of the Rayleigh-Taylor structures is seen there in projection. 
After summation over all the pixels, the total differential emissivity of the octant of SNR simulated is of the order of $10^{39}$~ph.s$^{-1}$.eV$^{-1}$ (that is a differential flux of the order of $10^{-5}$~ph.cm$^{-2}$.s$^{-1}$.eV$^{-1}$ at a distance of $1$~kpc). After summation over the broad energy band considered, the total integral emissivity of the octant is of the order of $10^{43}$~ph.s$^{-1}$ (that is an integral flux of the order of $10^{-1}$~ph.cm$^{-2}$.s$^{-1}$ at a distance of $1$~kpc). 

\flag{The effect of acceleration} can be seen on both the intensity levels and the patterns of the emissivity. 
The emissivity of the shocked ISM is always low (compared to the ejecta), with back-reaction at the forward shock it is even lower, plus the emitting region is substantially narrower. The emissivity of the ejecta remains about the same, as a result the contrast between the ejecta and the ISM becomes more pronounced around the contact discontinuity. And as the total emission is dominated by the ejecta, it is not much affected by particle back-reaction (by about 18\% in the case simulated here). We also note that the general aspect of the ejecta is not visibly modified, although the Rayleigh-Taylor structures at small scales are all spatially re-distributed (a quantification of the degree of turbulence would be interesting, but is beyond the scope of this paper). 

\strong{Spectra by medium}. To discuss more precisely the impact of acceleration on the ISM and the ejecta, on Figure~\ref{fig:spc_map} we plot the broad-band energy spectra spatially integrated over different regions. The top black curves show the emissivity of the entire region simulated -- this offers an orthogonal view to Figure~\ref{fig:prj_THtt_broad}: instead of summing all energy bins in each pixel, here we sum all pixels in each energy bin. The blue (red) curves show the emissivity of a region dominated by the ISM (ejecta), defined as all the (projected) cells where the fraction of ISM (ejecta) is $\geq 95\%$. We already noted that the emissivity of the ISM is substantially lower than the emissivity of the ejecta. Now it becomes apparent that the spread increases with energy. On the plot, we have indicated a few energy bands that are dominated by a particular emission line of a unique element -- many more lines can be seen in the spectrum, but they blend with each other at the spectral resolution chosen. The emissivity of a line depends on the abundances of the elements, which are different in the ISM and in the ejecta (but which get mixed in the vicinity of the contact discontinuity). For instance, one can see prominent lines of Silicon and Sulfur in the ejecta. For a given element, the emissivity of the lines also depends on its ionization state, which in turn depends on the density, temperature, and age of the cell, and therefore varies strongly with position (see next section). As a result, spectra with distinctively different shapes (not shown here) can be obtained when extracting small spatial regions in the projected maps. 

\flag{The effect of acceleration} is to reduce the intensity of the emission, all the more so at the highest photon energies. The situation is very different for the ISM and for the ejecta,\footnote{Here we warn the reader, before making direct comparisons between the spectra of the two media and in the two acceleration cases, that the number of extracted cells varies, as indicated in the legend of the figure.} since we consider efficient acceleration to occur at the forward shock. For the ISM-dominated region, with back-reaction the emissivity is reduced by a factor of about~2 around 1~keV and by more than 2~orders of magnitude at 10~keV. This will have a strong impact on the visibility of the post-shock regions with an X-ray instrument. The total emissivity of the SNR is always dominated by the ejecta, though, and so is less affected by back-reaction: it is reduced by about 30\% at 100~eV and about 75\% at 10~keV. As a result, the relative visibility of the ISM and of the ejecta is strongly affected by particle acceleration: back-reaction enhances the contrast between the ejecta and the ISM.
The drastic reduction of the emission of the ISM at high energies is explained by the exponential decrease of the continuum emission with temperature: the continuum emission dominates at high energies, and the temperature is much lower with back-reaction (see  Section~\ref{sec:results_diag_thermo}). We note that line emission is not affected in the same way: in fact, the few lines present at high energies become more prominent above the continuum. 

\strong{Spectra by Elements}. To conclude this section on the broad-band emission, we decompose the total spectrum of Figure~\ref{fig:spc_all} in another way, by separating the emission of the various elements handled by the code. We do not show all 15 here, but only 3 of particular importance: Oxygen, Silicon, and Iron on Figure~\ref{fig:spc_all}. As the atomic number increases, lines appear at higher and higher energies (provided the conditions are met in the plasma). 

\flag{The effect of acceleration} is fairly subtle when isolating a single element in this way. The emission is always lower with back-reaction, but the differences are small, and mostly seen at the highest energies (at 10~keV, they reach 40-50\%). We are showing here three heavy elements, that are more abundant in the ejecta. Similar behaviour is obtained for the other elements.
We must emphasize here that the spectra plotted on Figure~\ref{fig:spc_all} are spatially-integrated spectra, that show the global properties of the remnant. Actually the emission varies a lot spatially because of the different conditions in density and temperature, and the back-reaction of particles does affect the local conditions, so that spectra for any element can have distinctively different shapes when looking at specific regions of the SNR. 

\subsubsection{Emission in narrow energy bands}
\label{sec:results_emission_narrow}

As the former analysis showed that most features are energy-dependent, we now present emission maps for three of the narrow energy bands displayed on Figure~\ref{fig:spc_map} and~\ref{fig:spc_all}: the `O-K band' ($545-580$~eV) on Figure~\ref{fig:prj_THtt_OK}, the `Si-K band' ($1715-1840$~eV) on Figure~\ref{fig:prj_THtt_SiK}, and the `Fe-K band' ($6250-6520$~eV) on Figure~\ref{fig:prj_THtt_FeK}. These bands were chosen from the presence of a prominent line of an element (for the conditions of this particular remnant at this particular age) and labeled with the name of this element, however they are defined by a range of energies, and will invariably contain a continuum emission from all elements. Because they are narrow bands centered at different energies, they probe different conditions of the plasma, and therefore different regions of the SNR, and different levels of the impact of particle acceleration. To make this even more apparent, we propose another kind of visualization on Figure~\ref{fig:prj_THtt_rgb}: RGB composites made out of the emissivities in the O-K, Si-K and Fe-K bands (corresponding to  low, medium, and high photon energies, respectively). 

\strong{Line emissivity maps.} At the top of the figure for each band, as well as on the composite figure, we show the total emissivity in the band (as on Figure~\ref{fig:prj_THtt_broad}). Note that the intensity scale is different on each figure, because it varies by 6 orders of magnitude. 
The turbulent pattern of the ejecta is visible on all emissivity maps, but with notable variations. The O-K and Si-K bands probe the deepest parts of the ejecta (i.e., the feet of Rayleigh-Taylor fingers), while the Fe-K band probes the shallowest parts of the ejecta (i.e., the mushroom heads of Rayleigh-Taylor fingers). There is no radial stratification of the different elements in our initial setup, so what is observed here is the effect of the variations in the plasma conditions. This is caused by the fact that the Fe-K line requires high temperatures, which are found only close to the contact discontinuity, while the O-K and Si-K lines shine at pretty much all temperatures found in the shocked region, including close to the reverse shock (model emissivities in the four bands chosen, as a function of temperature and ionization age, are provided for reference in Appendix~\ref{sec:app_linem}). Hence observing the remnant at different energies, corresponding to different elements, allows to perform a radiography of its structure. 

\flag{The effect of acceleration} is not to be seen on the global pattern of emission, but rather on the local variations in emission: the distribution of Rayleigh-Taylor structures is slightly modified with back-reaction. As before, the impact is more pronounced at high energies: variations are more visible in the Fe-K band. Because the temperature is lower, with back-reaction the top of the Rayleigh-Taylor fingers is no longer bright, and the Fe-K map gets closer to the maps of other elements. These effects are not however very prominent on the emissivity maps because the emission is proportional to the square of the density, and the density of the ejecta (which make most of the emission in terms of intensity and patterns) is not strongly affected by acceleration at the forward shock.

\strong{Equivalent width maps.} To focus on the line emissivity, we now present maps of `equivalent width' at the bottom of Figures~\ref{fig:prj_THtt_OK}, \ref{fig:prj_THtt_SiK}, \ref{fig:prj_THtt_FeK}. These maps are produced by first subtracting the continuum emission from the total emission (in the energy band selected) to get only the emission of the line, then by dividing by the continuum emission to remove the strong dependence of the emission on density \citep{Hwang2000a, Miceli2006a}. The value obtained represents the width (in energy) of the underlying continuum that one needs to consider to obtain the same number of photons as in the line. The dependence on density being removed, this value is now only a function of the electronic temperature, of the abundance of elements (which we chose to be uniform in each medium), and on the ionization state of each element (which depends on density and temperature). It is remarkable that the maps obtained are not uniform: the ejecta pattern is still clearly visible (although it looks different than before). 

\flag{The effect of acceleration} is much more visible on the equivalent width maps than on the emissivity maps. As before, back-reaction effects are more pronounced at higher energies: they are limited in the O-K band, but important in the Fe-K band. The effects are two-fold. First, the maximum width gets higher (by about 75\% in the Fe-K band), which means that back-reaction tends to enhance the line emission with respect to the continuum emission. This is caused by the fact that the line and continuum emission have a different dependence to the changes in temperature induced by particle acceleration. The main effect of back-reaction is to lower the temperature in the ISM, and therefore to lower its emission. Given that the ISM emission is mostly a continuum emission, while most of the line emission comes from the ejecta, the result is that the diffuse ISM emission is removed and the emission lines of the heavy ejecta become more evident. Secondly, the contrast decreases in the interior region of the SNR (by about 25\% in the Fe-K band), which means that back-reaction tends to smoothen the line emissivity spatially. This is partly due to the fact that the shocked region shrinks, so that all features get compressed.

\strong{Continuum emissivity maps.} To conclude this section, we show on Figure~\ref{fig:prj_THtt_cont} emission maps in two broader energy bands, located at high energies: the `cont1' band from 4~keV to 6~keV and the `cont2' band from 8~keV to 10~keV. In our model these bands are devoid of line emission (note that we do not include the faint contributions from Chrome and Manganese), and therefore illustrate the fate of the continuum emission. The continuum emission directly scales with density, so that these maps are mostly maps of the (projected) density squared. The continuum emission also decreases exponentially at high temperatures, so that the maps are not exactly the same in the two energy bands.

\flag{The effect of acceleration} on the shocked ISM is highlighted, because we have excluded emission lines which mostly trace the ejecta. As particles are accelerated at the forward shock, the temperature of the ISM is significantly reduced downstream of shock, and thus its high energy emission. This can be seen on the projected maps when looking towards the edges of the remnant, between the contact discontinuity and the forward shock: without back-reaction the emissivity is low immediately downstream of the shock and rises when getting closer to the ejecta, with back-reaction the shock front is closer from the ejecta, but the emissivity remains low everywhere downstream of it. This can also seen when looking towards the interior of the remnant (where the ISM is on top of the ejecta): with back-reaction, a veil of diffuse emission from the ISM is removed. 


\section{Discussion}
\label{sec:discussion}

In this section we summarize the most salient features we have observed on the various simulated maps and spectra, and how they impact the interpretation of observations. We~also review the assumptions underlying our model, to assess how the results might differ with a different setup.

\subsection{General morphology and energy transfers}
\label{sec:discussion-morphology-energy}

Even before conducting a detailed computation of the thermal emission, one expects morphological modifications to be visible, that are caused by the presence (and escape) of a significant population of relativistic particles. In the simulation presented here, we estimate that particle acceleration took away about 20\% of the kinetic energy released by the explosion. 
As the resulting fluid gets more compressible, the shocked region shrinks \citep{Decourchelle2000a}: it gets narrower and denser, as seen on Figure~\ref{fig:cut_diag_hydro}. The first effect: a more compact structure can provide an observational diagnostic, as already shown in \cite{Ferrand2010a} and associated references. The second effect: a higher compression ratio is more difficult to measure, because it requires the detection of the thermal emission from the plasma located just behind the shock, a region which was found to be usually dominated by non-thermal (synchrotron) radiation.

\begin{figure}
\renewcommand{\figwidth}{16cm}
\centering
\includegraphics[width=\figwidth]{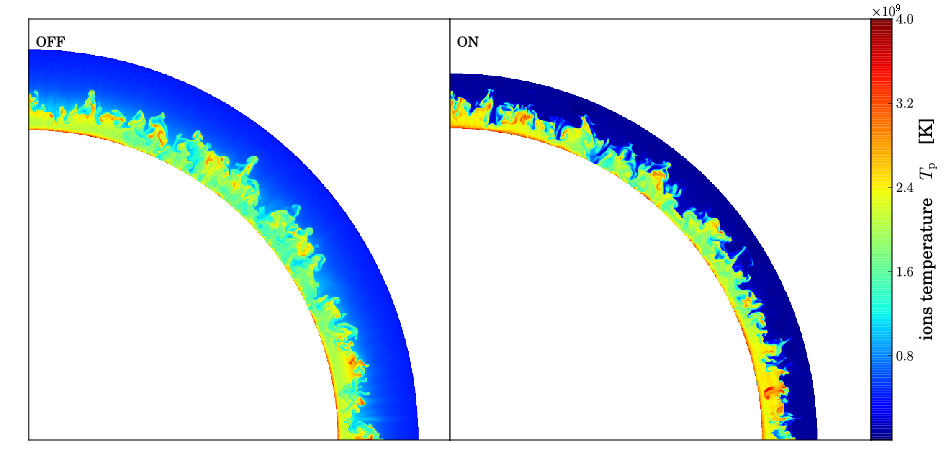}
\includegraphics[width=\figwidth]{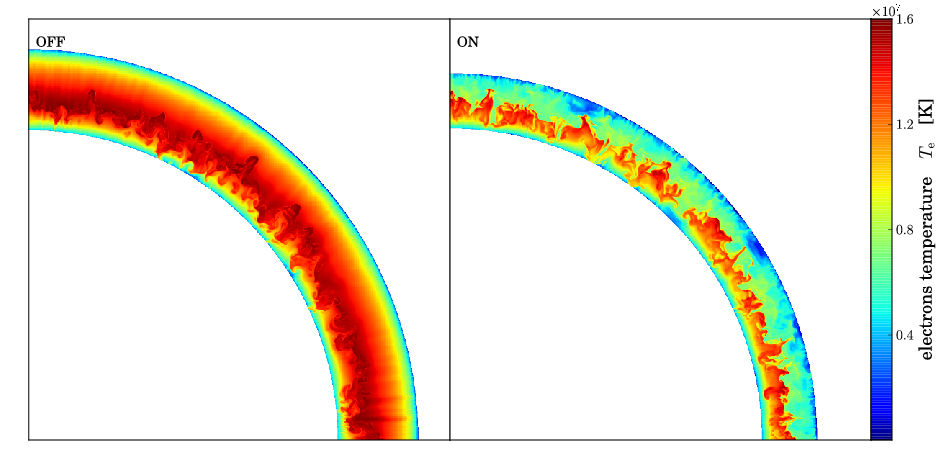}
\caption{Slices of the temperature, in the $z=0$ plane, in the shocked region.
Top: protons, bottom: electrons.
Cases without (`OFF', on the left) and with (`ON', on the right) back-reaction of particles are compared.
\label{fig:cut_diag_T}}
\end{figure}

\begin{figure}
\renewcommand{\figwidth}{16cm}
\centering
\includegraphics[width=\figwidth]{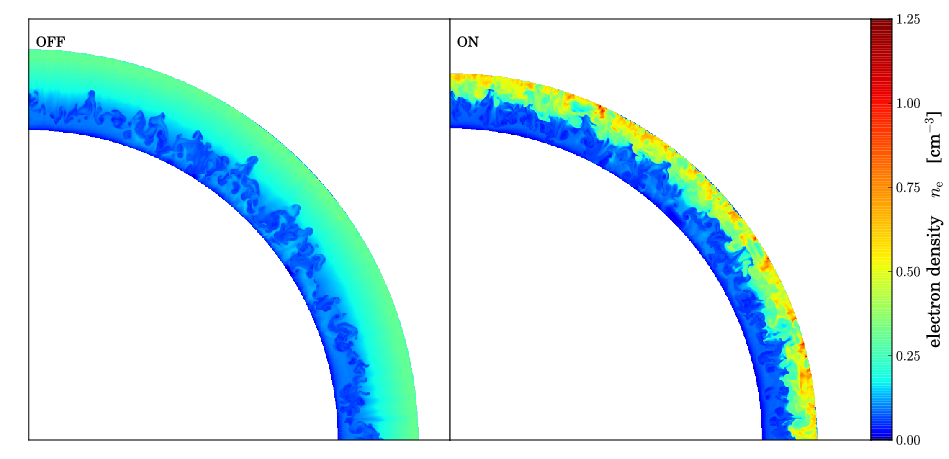}
\caption{Slices of the electronic density, in the $z=0$ plane, in the shocked region.
Cases without (`OFF', on the left) and with (`ON', on the right) back-reaction of particles are compared.
\label{fig:cut_diag_ne}}
\end{figure}

\begin{figure}
\renewcommand{\figwidth}{14cm}
\centering
\includegraphics[width=\figwidth]{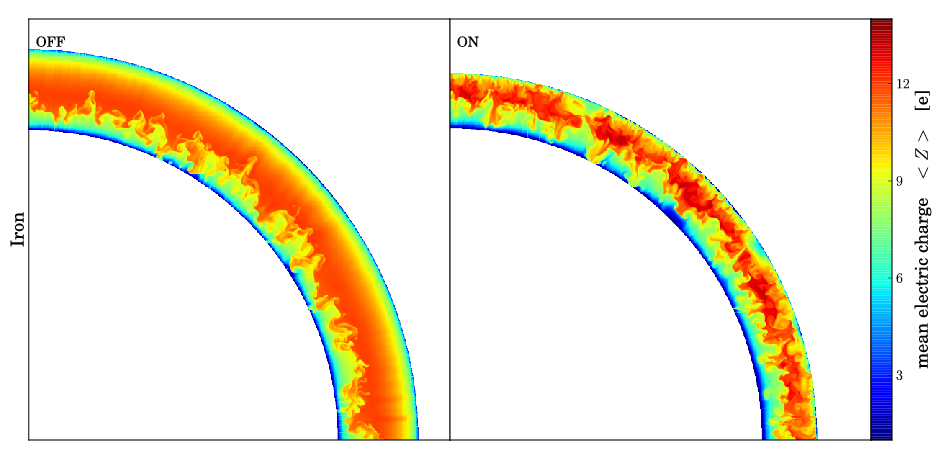}
\includegraphics[width=\figwidth]{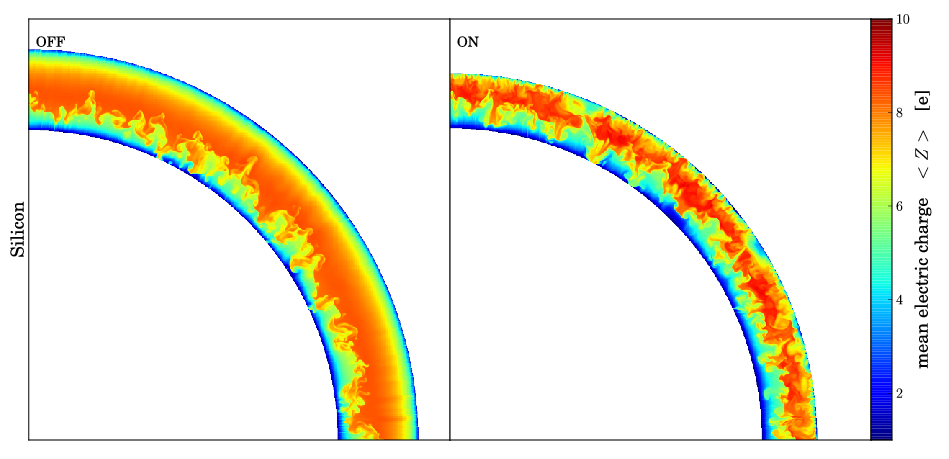}
\includegraphics[width=\figwidth]{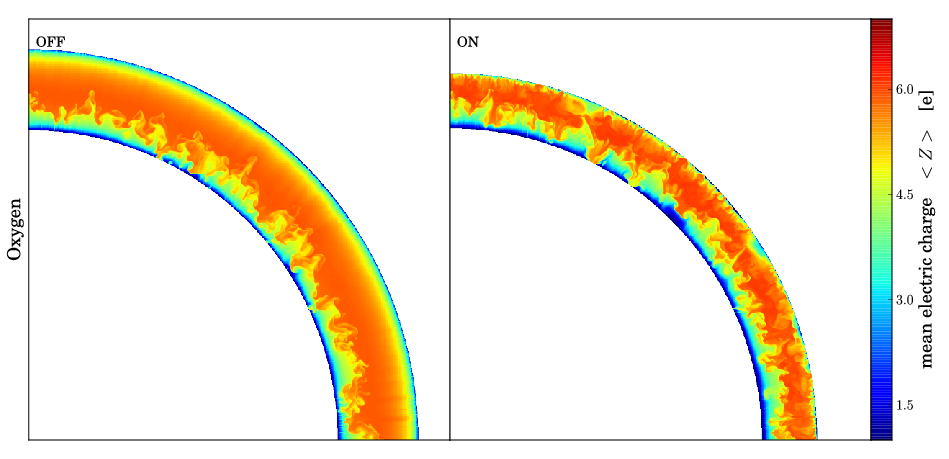}
\caption{Slices of the mean electric charge of a given element, in the $z=0$ plane, in the shocked region.
Top: Iron, middle: Silicon, bottom: Oxygen.
Cases without (`OFF', on the left) and with (`ON', on the right) back-reaction of particles are compared.
\label{fig:cut_diag_Z}}
\end{figure}

\begin{figure}
\renewcommand{\figwidth}{16cm}
\centering
\includegraphics[width=\figwidth]{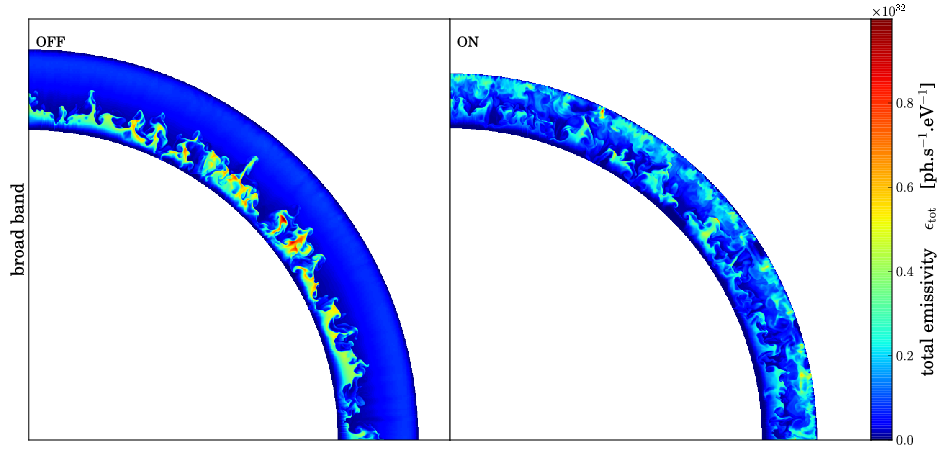}
\includegraphics[width=\figwidth]{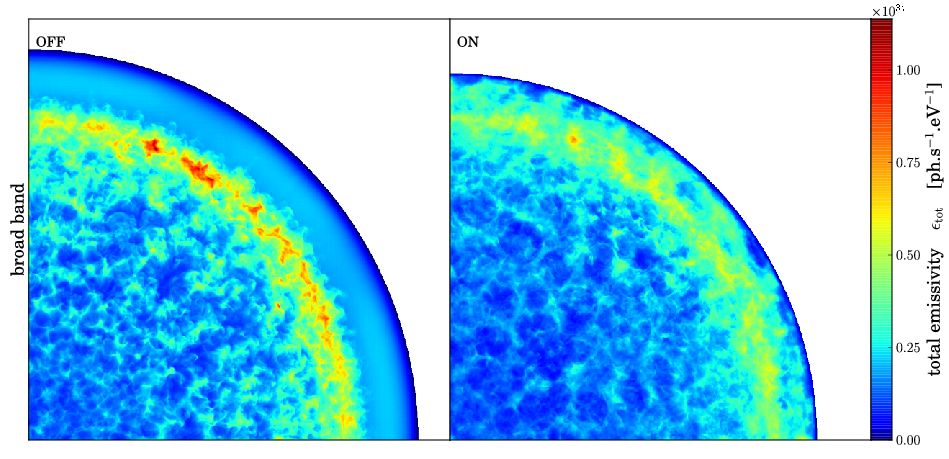}
\caption{Total thermal emissivity in a broad $0.3-10$~keV band.
Top: slices in the $z=0$ plane, bottom: projected maps along the $z-$axis.
On both types of maps, the differential emissivity at the source $\epsilon_{\rm tot}$ is plotted for each simulated pixel. The flux at the Earth will be diluted according to the square of the distance~$d$ to the source. The actual fluxes recorded by a given instrument will depend on its own angular resolution (we recall that the maps presented here have $1024$ pixels along each direction, at the age shown here they span $5.61$~pc, hence a physical cell size of $\simeq 0.0055$~pc, and an angular resolution of about $1"$ at $d=1$~kpc). 
Cases without (`OFF', on the left) and with (`ON', on the right) back-reaction of particles are compared.
\label{fig:prj_THtt_broad}}
\end{figure}

Besides the density, the level of thermal emission critically depends on the temperature of the plasma, more precisely on the electrons temperature ($T_e$). The unavoidable effect of efficient particle acceleration at the shock is a decrease of the temperature of the shocked material. The drop is large for ions (Figure~\ref{fig:cut_diag_T}), which take part in the dynamics of the shock: in the case presented $T_p$ is reduced by up to one order of magnitude in the shocked ISM. However, the situation is different for electrons: there is no reason why they would be heated in the same way as protons at the shock, and their temperature will also depend on subsequent energy exchanges downstream of the shock. Assuming, as we do, a progressive Coulombian equilibration between electrons and protons, $T_e$ appears to be always limited by the age of the remnant. $T_e$ is still affected by particle acceleration, indirectly, through the decrease of $T_p$: it is lowered by a factor of up to about~2 in the shocked ISM (Figure~\ref{fig:cut_diag_T}). If~we were considering the limit case of instant equilibration of electrons with ions at the forward shock, then $T_e$ would be higher than in our model, but the effect of back-reaction would be even more drastic. We~note that the obtained values of $T_e$ are in a range where  emission lines can be very sensitive (see Figure~\ref{fig:linem}), and that the continuum emission has an exponential dependence on $T_e$, so that even small variations of $T_e$ can have a major impact on the thermal emission. 
\\

\subsection{Energy-dependent effects: the need for spectroscopy}
\label{sec:discussion-spectral}

\begin{figure}
\renewcommand{\figwidth}{16cm}
\centering
\includegraphics[width=\figwidth]{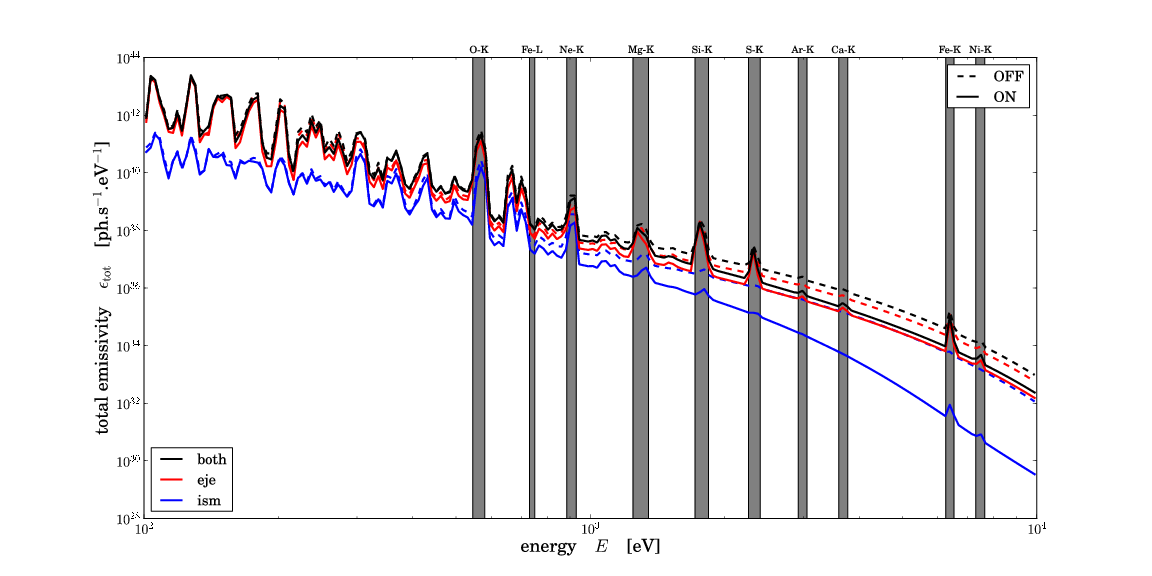}
\caption{Broad-band energy spectra, spatially integrated over different regions, with or without the back-reaction of particles (solid or dashed lines). 
Top curves (in black) show the sum of the emissions from all cells (note that there are 12\% less emitting cells in the ON case with respect to the OFF case).
Middle curves (in red) show the contribution of cells dominated by the shocked ejecta, which make 34\% of all the emitting cells in the un-modified case (OFF), and 29\% of the cells in the modified case (ON).
Bottom curves (in blue) show the contribution of cells dominated by the shocked ISM, which make 13\% of all the emitting cells in the un-modified case (OFF), and only 6\% of the cells in the modified case (ON).
Some prominent lines are labeled by the emitting element for reference.
\label{fig:spc_map}}
\end{figure}

\begin{figure}
\renewcommand{\figwidth}{16cm}
\centering
\includegraphics[width=\figwidth]{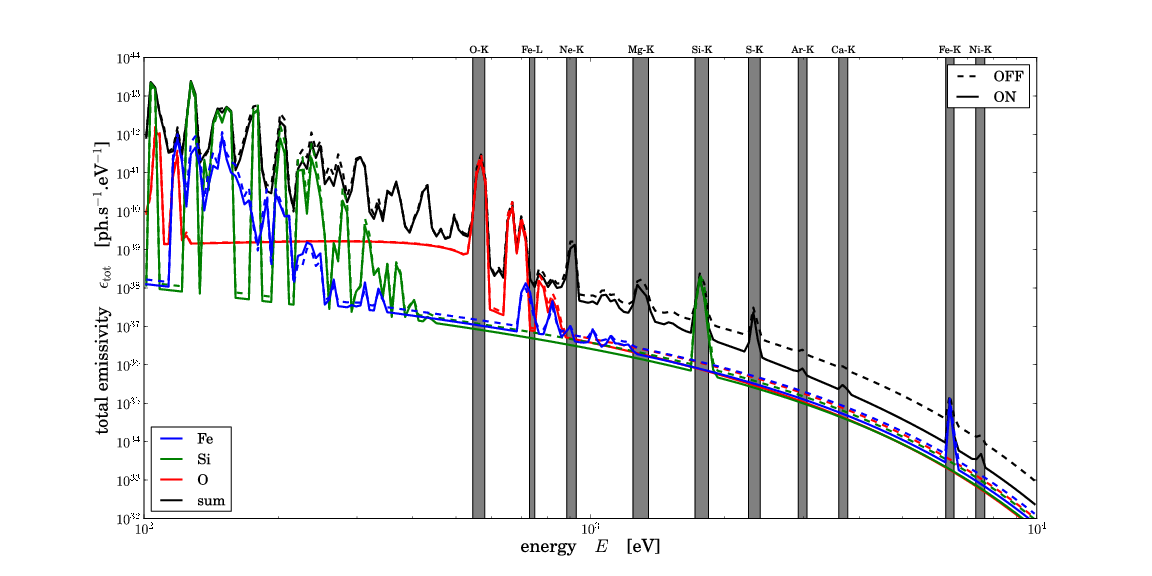}
\caption{Broad-band energy spectra for three elements (in colours: Oxygen in red, Silicon in Green, Iron in Blue) and sum of the contributions of all elements considered (in black), with or without the back-reaction of particles (solid or dashed lines), integrated over the whole octant of the remnant simulated. Some prominent lines are labeled by the emitting element for reference.
\label{fig:spc_all}}
\end{figure}

\begin{figure}
\renewcommand{\figwidth}{16cm}
\centering
\includegraphics[width=\figwidth]{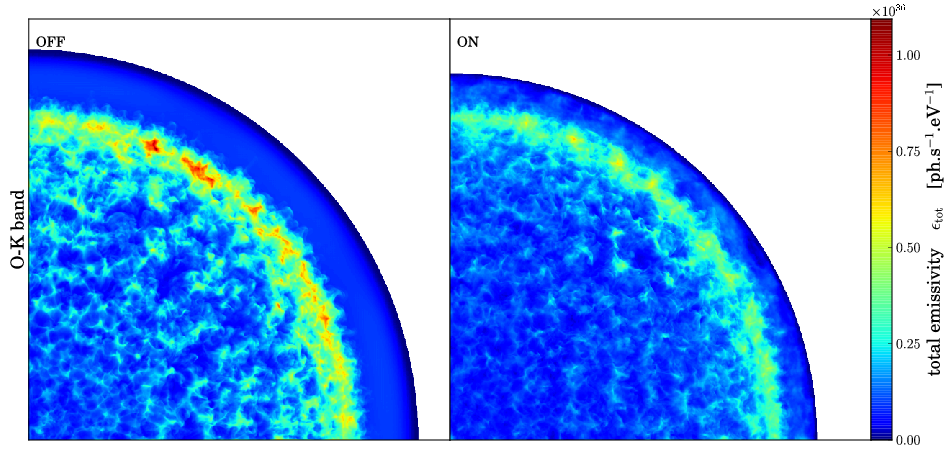}
\includegraphics[width=\figwidth]{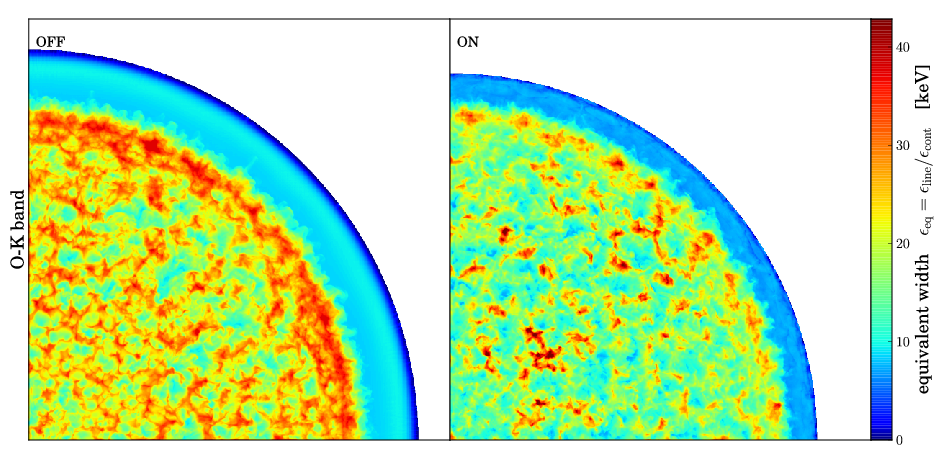}
\caption{Projected maps, along the $z-$axis, of the thermal emission in the `O-K band' ($545-580$~eV).
Top: emissivity, bottom: ratio of lines over continuum emissivity (equivalent width maps).
Cases without (`OFF', on the left) and with (`ON', on the right) back-reaction of particles are compared.
\label{fig:prj_THtt_OK}}
\end{figure}

\begin{figure}
\renewcommand{\figwidth}{16cm}
\centering
\includegraphics[width=\figwidth]{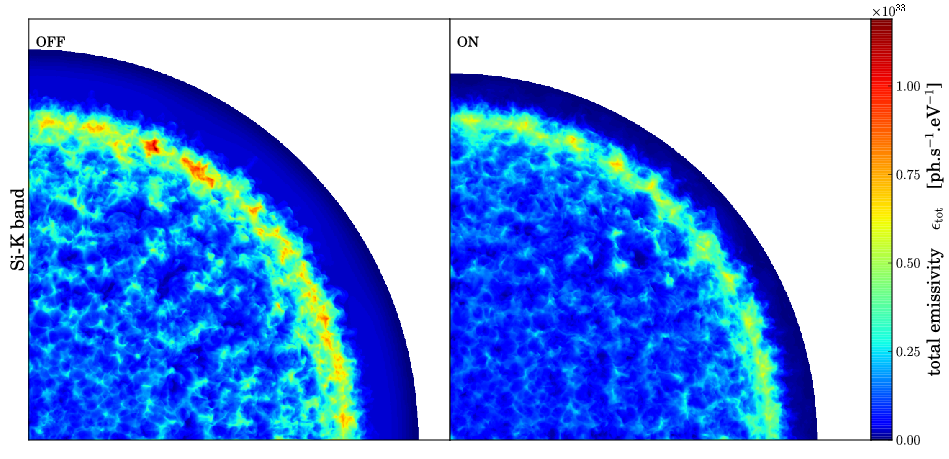}
\includegraphics[width=\figwidth]{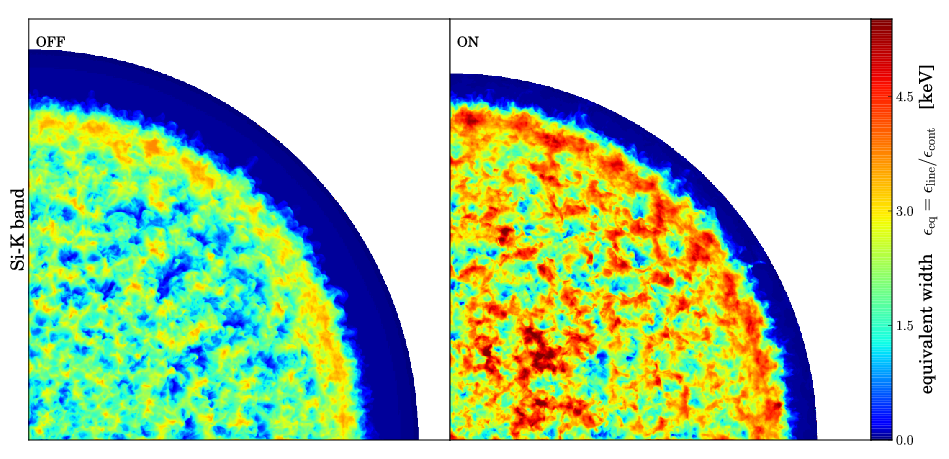}
\caption{Projected maps, along the $z-$axis, of the thermal emission in the `Si-K band' ($1715-1840$~eV).
Top: emissivity, bottom: ratio of lines over continuum emissivity (equivalent width maps).
Cases without (`OFF', on the left) and with (`ON', on the right) back-reaction of particles are compared.
\label{fig:prj_THtt_SiK}}
\end{figure}

\begin{figure}
\renewcommand{\figwidth}{16cm}
\centering
\includegraphics[width=\figwidth]{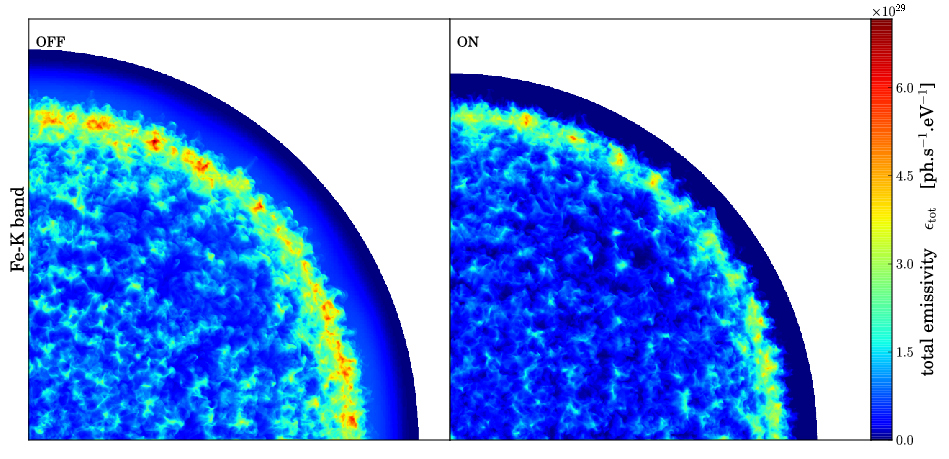}
\includegraphics[width=\figwidth]{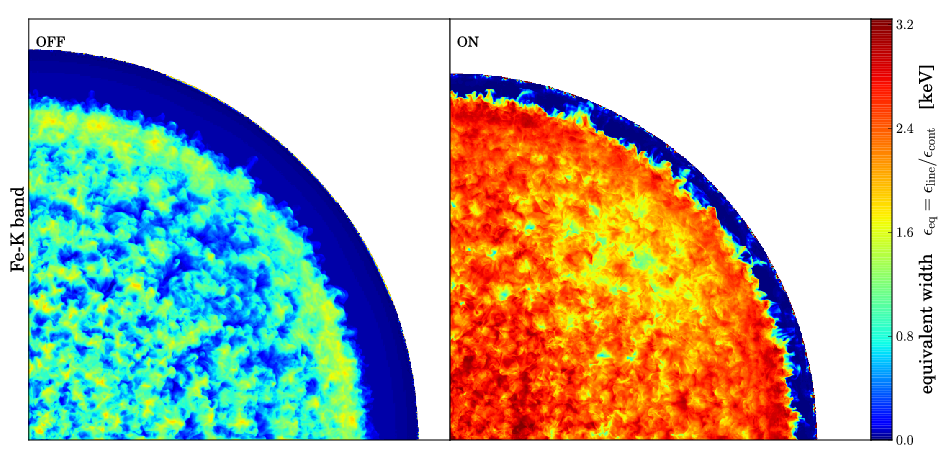}
\caption{Projected maps, along the $z-$axis, of the thermal emission in the `Fe-K band' ($6250-6520$~eV).
Top: emissivity, bottom: ratio of lines over continuum emissivity (equivalent width maps).
Cases without (`OFF', on the left) and with (`ON', on the right) back-reaction of particles are compared.
\label{fig:prj_THtt_FeK}}
\end{figure}

\begin{figure}
\renewcommand{\figwidth}{16cm}
\centering
\includegraphics[width=\figwidth]{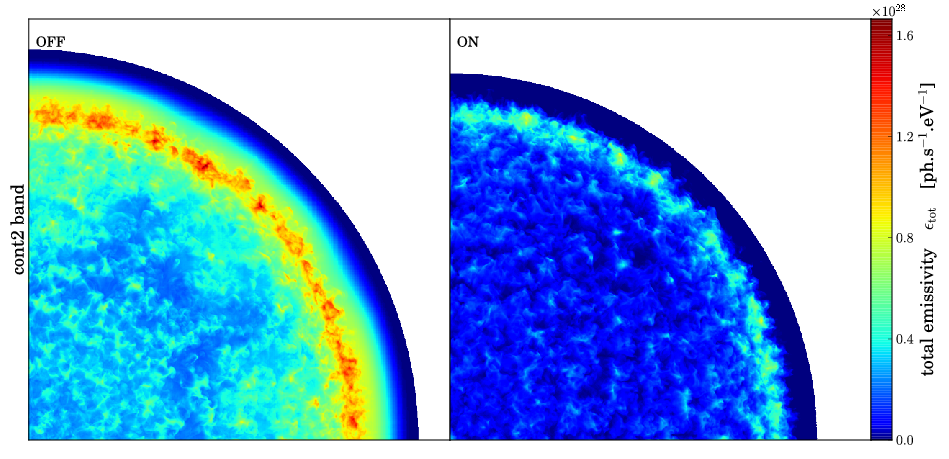}
\includegraphics[width=\figwidth]{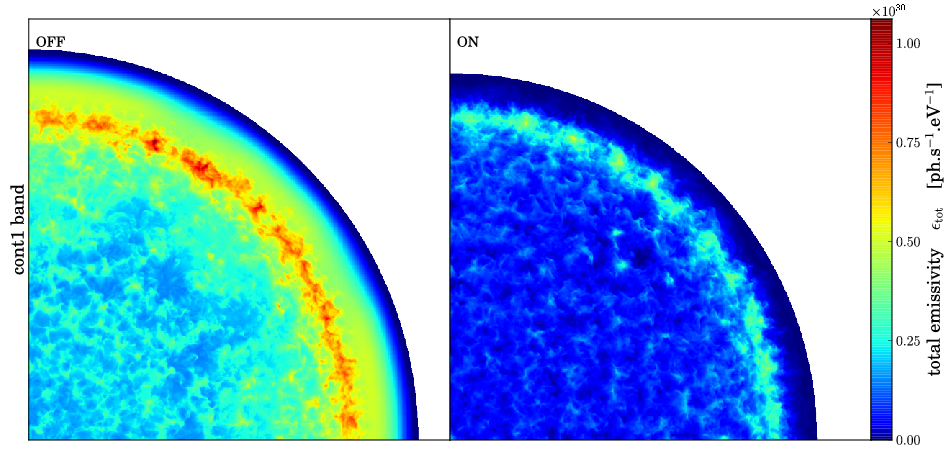}
\caption{Projected maps, along the $z-$axis, of the thermal emission flux in two high-energy continuum bands. 
Top: $8-10$~keV, bottom: $4-6$~keV.
Cases without (`OFF', on the left) and with (`ON', on the right) back-reaction of particles are compared.
\label{fig:prj_THtt_cont}}
\end{figure}

\begin{figure}
\renewcommand{\figwidth}{16cm}
\centering
\includegraphics[width=\figwidth]{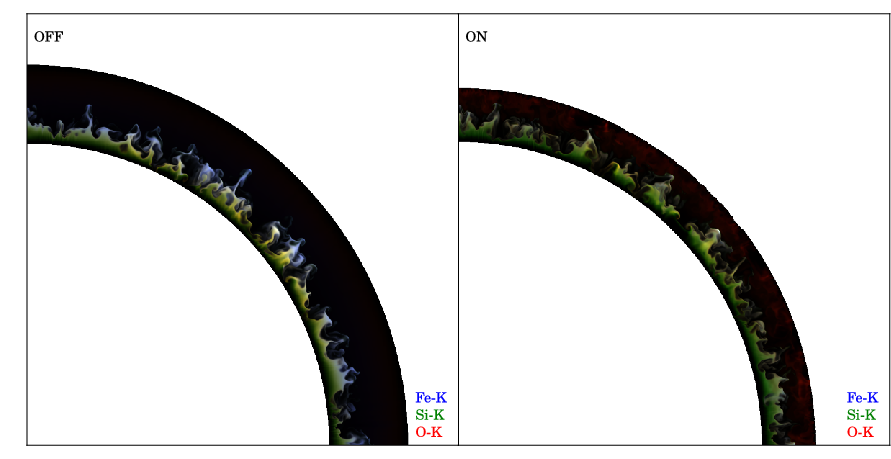}
\includegraphics[width=\figwidth]{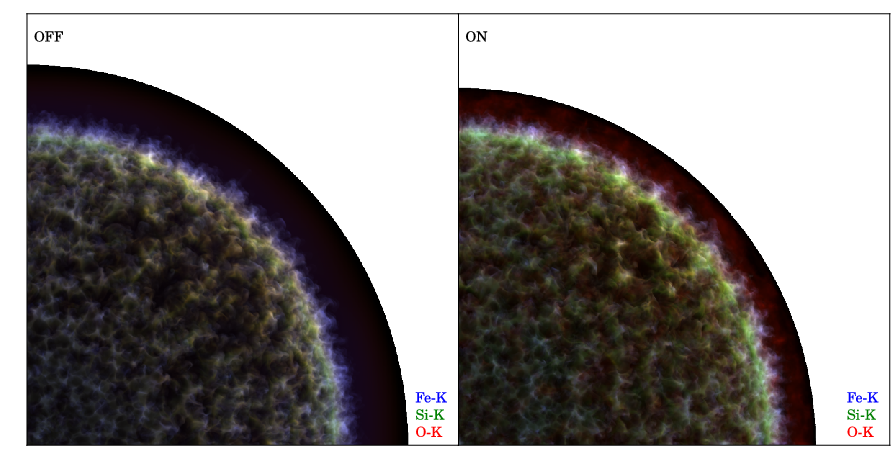}
\caption{RGB rendering of three line emissions. For each pixel, the value of the red / green / blue channel is assigned from the emissivity of resp. the O-K band / Si-K band / Fe-K band, as displayed on Figures \ref{fig:prj_THtt_OK} / \ref{fig:prj_THtt_SiK} / \ref{fig:prj_THtt_FeK} (linearly normalized to 256 levels). Regions of pure blue, for instance, are dominated by Fe-K line emission. Regions of yellow = red + green, are made out of a blend of O-K and Si-K line emission.
Top: slices in the $z=0$ plane, bottom: projected maps along the $z-$axis.
Cases without (`OFF', on the left) and with (`ON', on the right) back-reaction of particles are compared.
\label{fig:prj_THtt_rgb}}
\end{figure}

We always observed that, as part of the energy is channeled into energetic particles at the shock, the thermal emission of the plasma is reduced. However, the overall impact is small (less than 10\% in our simulation), and variations in the absolute value of the X-ray flux are unlikely to be used as signatures of particle acceleration given the uncertainties on all the relevant parameters. Much more reliable will be the comparison of fluxes in different energy bands. Indeed, we observed a strong energy dependence of back-reaction effects, as can be seen on spatially-averaged spectra (Figure~\ref{fig:spc_map}) or by comparing maps in different energy bands (Figures \ref{fig:prj_THtt_OK}, \ref{fig:prj_THtt_SiK}, \ref{fig:prj_THtt_FeK}, \ref{fig:prj_THtt_cont}). As a rule, the higher the energy of the photons observed, the higher the reduction in the emissivity. As a result, observing the remnant at different energies probes different levels of impact of the back-reaction of particles on the shocked material. In general effects can be measured only above $\sim$1~keV, and they are much more obvious for the ISM. Despite the increase in density downstream of the forward shock, we find that the high-energy emission from this region is strongly reduced. This is due to the fact that the ISM emission is mostly a continuum emission, which has a strong dependence on temperature. Emission lines have a different dependence on temperature, and so are not affected in the same way: actually the reduction of the continuum caused by back-reaction makes the lines more prominent (this can be seen on the spectrum of Figure~\ref{fig:spc_map} for the ISM as well as on the equivalent width maps of Figure~\ref{fig:prj_THtt_FeK} for the ejecta). 

\subsection{Space-dependent effects: the need for spatially-resolved spectroscopy}
\label{sec:discussion-spatial}

For a given element, the effects of back-reaction are barely visible on overall spectra (Figure~\ref{fig:spc_all}), where the emission is spatially-averaged over the whole SNR. But on the emission maps, spectra can look noticeably different from one pixel to another, and in the same location when considering back-reaction or not. In particular, it is important to separate the ISM and the ejecta, which have different emission properties, and are affected in a different way by back-reaction, but which are mixed up by Rayleigh-Taylor instabilities. Regarding simulations, this implies a 3D treatment to get the turbulent distribution of the shocked material. Regarding observations, this is a call for spatially-resolved spectroscopy. Given the strong variations in density and temperature that we observe on the maps, the \emph{average} quantities that are commonly\footnote{The observer can also perform multi-component fitting, see e.g.~\cite{Kosenko2010c}.} derived from fits of the global spectra may be an oversimplification. For instance, in which sense the temperature derived from the spatially-averaged emission can be regarded as a spatial average of the temperature of all the emitting cells should be clarified. Even when a 2D spatially-resolved study is conducted, we recall the obvious fact the emission is integrated along the line-of-sight. Only 3D simulations can properly explain how the resulting features, as seen in projection with modern spectro-imaging instruments, are related to the local plasma conditions -- as affected by particle acceleration.

The spatial effects of back-reaction are much better seen on equivalent width maps rather than on total emissivity maps, after pure density effects have been removed, in order to reveal temperature and abundances effects on ionization and emission. We~note that our synthetic equivalent width maps are not uniform at all, despite the use of uniform abundances (respectively in the ISM and in the ejecta): they map variations in the plasma conditions (temperature and ionization age), which have a non-trivial pattern in 3D (and even more when seen in projection) because of the hydrodynamic instabilities. So observing these could not be directly used to map the abundances of the elements, at least on small scales. If~the various elements were not distributed evenly spatially, one could see bigger clumps of emission on the maps, as observed in some remnants. We deliberately kept the description of the ejecta simple in this paper, and regard the simulations presented here as a reference case. The next step will be to initialize the simulations with a non-uniform distribution of the ejecta, as obtained from supernova models, to reflect the combustion layers formed in the progenitor stars and/or during the explosion (see for instance~\cite{Decourchelle2001a} for observational evidence for incomplete mixing of the Silicon and Iron layers). This can be done in our current numerical setup by the addition of a passive tracer for each species we want to follow.

\subsection{The fate of the ejecta}
\label{sec:discussion-ejecta}

The impact of back-reaction of accelerated particles is distinctly different for the (shocked) ISM and for the (shocked) ejecta. This of course primarily stems from the fact that we consider acceleration at the forward shock only: the temperature of the ISM is much lowered, whereas the ejecta are only affected indirectly, through the dynamics around the contact discontinuity. Another aspect is that the two media are not affected in the same way by changes in the plasma conditions, because of their different composition: most emission lines of heavy elements, present in the ejecta, are less affected by temperature than the continuum. As a result, the visibility of the ejecta with respect to the ISM is enhanced by the back-reaction. We note that indeed, on X-ray images of Tycho's SNR, the shocked ejecta shaped by Rayleigh-Taylor structures are clearly visible, whereas a shocked ISM component is not required to explain the observations \citep{Warren2005a}. On the other hand, on our simulated maps the contrast inside the ejecta seems to be somewhat reduced by the back-reaction (this is visible on the equivalent width maps of Figures \ref{fig:prj_THtt_OK}, \ref{fig:prj_THtt_SiK}, \ref{fig:prj_THtt_FeK}).

As the region between the contact discontinuity and the forward shock shrinks, the density and temperature profiles are modified on one side of the contact discontinuity, and the development of the Rayleigh-Taylor structures can be affected. In our earlier work \citep{Ferrand2010a} we did not observe a major impact on the overall growth of the fingers. In~this new study, looking at the X-ray emission in different energy bands allows to better see the effects of back-reaction on their structure. Radial variations of the line emissivities, caused by the evolution of temperature and ionization age downstream of the shocks, are visible on the RGB representations of Figure~\ref{fig:prj_THtt_rgb}. With back-reaction, the low-energy emission in the O-K and Si-K bands gets more prominent and the high-energy emission in the Fe-K band gets suppressed. So the morphology of the fingers is energy-dependent, and as a result is acceleration-dependent.
We note that the Fe-K emission observed with \textit{Chandra} in Tycho was used by \cite{Warren2005a} to trace the position of the reverse shock, whereas in our synthetic maps the Fe-K emission (Figure~\ref{fig:prj_THtt_FeK}) rather traces the location of the contact discontinuity, because only there sufficient electronic temperatures are reached for this high-energy line. \cite{Warren2005a} assumed the presence of hot Iron close to the reverse shock, based on the observation made with \textit{XMM-Newton}, that the Fe-K emission peaks inside the Fe-L emission \citep{Decourchelle2001a}. In our simulation, all Fe-L lines are faint. This is in agreement with the dependence on plasma conditions plotted on Figure~\ref{fig:linem}: Fe-L line emission requires higher ionization ages that are difficult to reach in the ejecta. As a result, the Fe-L lines are always blended with stronger K~lines from other elements (mostly Oxygen and Neon), except in a small energy band around $730-750$~eV (see spectra of Figures~\ref{fig:spc_map} and~\ref{fig:spc_all}); and even there, the emissivity of the line is of the same order as the emissivity of the continuum (mostly~Hydrogen and~Helium). Maps in this narrow energy band (not shown here) reveal that the Fe-L emission can indeed extend further than the Fe-K emission, although at flux levels that are not observationally relevant. To have a prominent Fe-K line emission closer to the reverse shock, we would need to somehow reverse the temperature profile between the reverse shock and the contact discontinuity. This could be caused either by a steeper initial density profiles in the ejecta, such as the exponential profile used in \cite{Dwarkadas2000a}, or by stronger electron heating at the reverse shock: the model of Tycho of \cite{Badenes2005a} works best for $\beta_{\rm S} \simeq 0.03$.

Here we recall that a simplification made in our numerical model is to decouple the computation of the ionization of the plasma from the computation of its hydrodynamic evolution: we record the evolution of the ionization age, and compute the actual ionization state at the end of the hydro simulation. The most accurate (but more numerically demanding) way to handle ionization would be to do the time integration of the balance equations for each species concurrently with the time integration of the conservation equations for the global fluid. A~comparison of the two approaches can be found in~\cite{Patnaude2010a} for the shocked ISM, based on~1D simulations of a remnant in the Sedov phase. It is shown that post-processing the ionization leads to slightly over-estimating the final thermal X-ray emission, especially at low photon energies and for high plasma densities. We observed that back-reaction effects are the highest at the highest energies, and we chose a fairly small ambient density, so we do not expect dramatic changes to our results regarding the relative effects of particle acceleration. We also note that the intensity of the ionization coupling depends on the fraction of massive elements, so it is not expected to be critical for the shocked ISM (where back-reaction effects are the strongest in our model, by construction). In the ejecta the situation is more complicated, this is why we loop over the computation of the ionization (see end of Section~\ref{sec:model_SNR_ionization}), in order to get consistent estimates of the final electron density and temperature. If we are to consider acceleration of particles at the reverse shock, the direct coupling of the ionization of at least the most abundant species should be investigated.

\subsection{Acceleration at the reverse shock?}
\label{sec:discussion-reverse-shock}

Acceleration of particles at the main blast wave of a SNR is well established, so it appears mandatory to explore the impact of the acceleration on the development of the shocked ISM. The situation is less clear for the reverse shock, that propagates backwards inside the ejecta. The major argument against considering acceleration at the reverse shock is that, during the early expansion phase of the SNR, the magnetic field frozen in the ejecta gets diluted orders of magnitude below the levels required to accelerate particles. Magnetic turbulence is a key component of the diffusive shock acceleration mechanism: it traps particles close to the shock front, where they can gain energy. However, accelerated particles are believed to be able to generate themselves the magnetic turbulence that scatter them, by triggering various instabilities upstream of the shock. It is therefore conceivable, if the acceleration process can start at a very early stage, that the magnetic field gets so much amplified that the process is then self-sustained. We refer the reader to \cite{Ellison2005a} for an investigation of efficient non-linear acceleration at the reverse shock, and to \cite{Telezhinsky2012a} for a recent study in the test-particle regime (using 1D hydrodynamic simulations for both works). 
\cite{Telezhinsky2012a} find that the reverse shock can make a significant contribution of the total spectrum in the early expansion phase (up to about 400--600~years, which corresponds to our final age). Even in the linear regime, total spectra never look like single-shock test-particle solutions, so that the question arises of making the distinction between modifications caused by non-linear effects at the forward shock and/or by an additional contribution from the reverse shock. Authors agree that the maximum energy reached by particles at the reverse shock is always lower than at the forward shock, so that we anticipate a lesser impact on the high-energy emission from the remnant. Still, recent radio and X-ray observations have been interpreted as the non-thermal emission from the reverse shock region in some SNRs: Cas A \citep{Gotthelf2001a,Helder2008a}, Kepler \citep{DeLaney2002a}, RCW 86 \citep{Rho2002a}, 1E 0102.2-7219 \citep{Sasaki2006a}. We are not aware of any such indications for Tycho, which was used as our reference object. In any case, it seems appropriate to add acceleration at the reverse shock as a second step in future work.

\subsection{SN Type}
\label{sec:discussion-SN-type}

Finally, we recall that we considered in this paper the remnant of a fiducial type~Ia supernova, hence the assumption of a uniform circumstellar medium. The assumption of perfect spherical symmetry might be the reason why we observe such a strong brightening at the edges of the shocked ejecta seen in projection, which is sharper than what is observed in remnants of the kind simulated such as Tycho's SNR. In reality, the medium in which the remnant expands will always exhibit some inhomogeneities. Such features will be specific to a particular remnant though, whereas we are trying here to exhibit generic features of the acceleration and back-reaction. 

In the case of the remnant of a core-collapse supernova, we would need to consider the radial density profile of the circumstellar medium, shaped by the wind of the massive progenitor. This can be accounted for in our model, using a power-law profile of index $s=2$ for the ISM. This substantially affects the hydrodynamic profiles in the shocked region, and therefore the thermal emission.


\section{Conclusion}
\label{sec:conclusion}

We have developed a 3D model that simulates the morphological and spectral evolution of a young SNR undergoing efficient particle acceleration. To~the best of our knowledge, this is the first time the thermal X-ray emission is computed when considering conjointly (i)~the 3D development of hydrodynamical instabilities (Rayleigh-Taylor) in the shocked region, (ii)~the time-dependent back-reaction of particles accelerated at the blast wave, and (iii)~the progressive temperature equilibration and non-equilibrium ionization of the plasma. 

Because we assumed efficient acceleration to take place at the forward shock only, back-reaction effects mostly impact the shocked ISM (an additional contribution on the shocked ejecta from the reverse shock might be useful to consider for some remnants). Our simulations reproduce the most well-known effects: the shell of shocked material gets narrower, denser, and of lower temperature. Note that the energetics of the blast wave directly affects the temperature of protons, whereas the emissivity of the hot plasma depends on the temperature of electrons. We assumed that electrons are not efficiently heated at the shock (which is supported by both observations and other simulations, and can be regarded as a conservative hypothesis), and that electrons progressively equilibrate with protons through Coulombian interactions downstream of the shock (equilibrium is never reached at the age considered). This complicates the discussion of the impact of shock acceleration on the thermal emission. We observed that the electron temperature in the shocked ISM is significantly reduced (here by a factor of about two). X-ray emission is still granted, but will be significantly affected by the efficiency of particle acceleration.

Because less energy goes into the thermal fluid, the effect of particle acceleration, if any, is generally to lower the overall thermal emission. However, absolute variations of fluxes are small, and unlikely to be used as diagnostics given all the uncertainties in the modelling of any given object. More useful are the facts that variations in flux depend very much on the energy band considered, and that spectral features vary spatially. This demonstrates the need for spatially-resolved high-resolution spectroscopy, and the usefulness of such simulations. In our model of a young SNR evolving in a homogeneous ISM, we observed that, the higher the photon energy, the more pronounced the effects of back-reaction: they are barely visible below 1~keV, and are easily measurable at 10~keV. The continuum and line fluxes are not affected in the same way, because of their different dependence on temperature. As a result, we observed that back-reaction enhances the visibility of the lines over the continuum. Equivalent width maps (obtained by subtracting and dividing the continuum emission) were shown to be the most useful visualization of the emission, because they probe the temperature and ionization age, and thus the level of back-reaction. Note that, even assuming uniform abundances, we obtain non-uniform equivalent width maps, a property that must be kept in mind when interpreting X-ray observations.

On all sorts of maps, the interior region of the SNR bears the signature of the instabilities developing between the shocked ISM and the shocked ejecta. Assuming the ejecta are of radially uniform abundances, the Rayleigh-Taylor fingers have a variable radial extent in X-rays, because of the global increase of temperature and ionization age towards the contact discontinuity, that reflect the history of shock conditions and the transport downstream of the shock. As a result, their global morphology in narrow energy bands is affected by particle acceleration.
A more qualitative study would be in order to assess how the small scale distribution of turbulent features is modified. 
Unsurprisingly, the level emissivity of the ejecta is not very much affected by efficient acceleration at the forward shock. However, the relative visibility of the ejecta is clearly enhanced by the back-reaction of particles on the shocked ISM, as the veil of diffuse emission from the ISM is removed. On the other hand, the contrast within the ejecta is reduced: back-reaction seems to smoothen interior regions of the maps. 

In this study, we used physical parameters most relevant for a type~Ia SNR. However our model is fairly general, and could be adapted for a type~II SNR. For simplicity we assumed uniform ejecta, and a uniform ISM (with different elemental abundances in each). This work should therefore serve as a reference case for future studies where we will assess the impact (i) of a more complex distribution of the ejecta stemming from the explosion and (ii) of an imprint of the progenitor on the ambient medium. Regarding the first aspect, \cite{Orlando2012a} have recently shown that ejecta clumping has a stronger impact on the shock structure than efficient acceleration in SN 1006. Some of our present choices were inspired by Tycho's SNR, but we did not try to make an actual fit of the emission of any given SNR at this point. Actually, we do not reproduce the peak of Iron line emission close to the reverse shock, as suggested by X-ray observations of Tycho. This prompts a more detailed study of the evolution of the electron temperature in the shocked ejecta. 
The purpose of this paper was first to outline the method and demonstrate its usefulness. We can now consider making a more precise modeling of some well observed historical remnants.
As such, the work presented here provides a framework for the interpretation of current SNR observations (with \textit{XMM-Newton}, \textit{Chandra}, \textit{Suzaku}), as well as for the preparation of observations
with future X-ray missions (in particular \textit{Astro-H}).

In this paper, we focused on the thermal emission from the SNR, as a way to diagnose the presence of energetic particles through back-reaction on the flow. The acceleration model we are using also computes the spectra of particles accelerated at the shock. In a follow-up paper, we will present results on the non-thermal emission produced by these energetic particles present inside the shell. We will thus obtain a complete, broad-band, view of the morphology and X-ray emission from young SNRs.

\newpage
\section*{Acknowledgements}

This work has been partially funded by the ACCELRSN project ANR-07-JCJC-0008 in France, and has been supported in Canada by the Natural Sciences and Engineering Research Council (NSERC), Canada's Foundation for Innovation (CFI), the Canada Research Chairs (CRC) program,  and the Canadian Institute for Theoretical Astrophysics (CITA).
The numerical simulations were performed on the Titane super-computer at CEA/CCRT (France) and on the Nuit super-computer at the University of Manitoba (Canada).

The authors thank the anonymous referee for his/her useful comments.
\\

\appendix


\section{Effect of resolution}
\label{sec:app_resolution}

\begin{figure}
\renewcommand{\figwidth}{16cm}
\centering
\includegraphics[width=\figwidth]{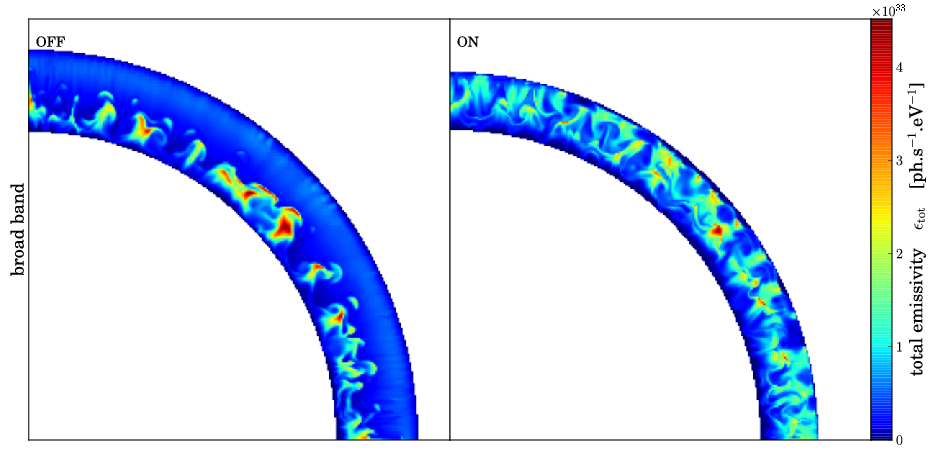}
\includegraphics[width=\figwidth]{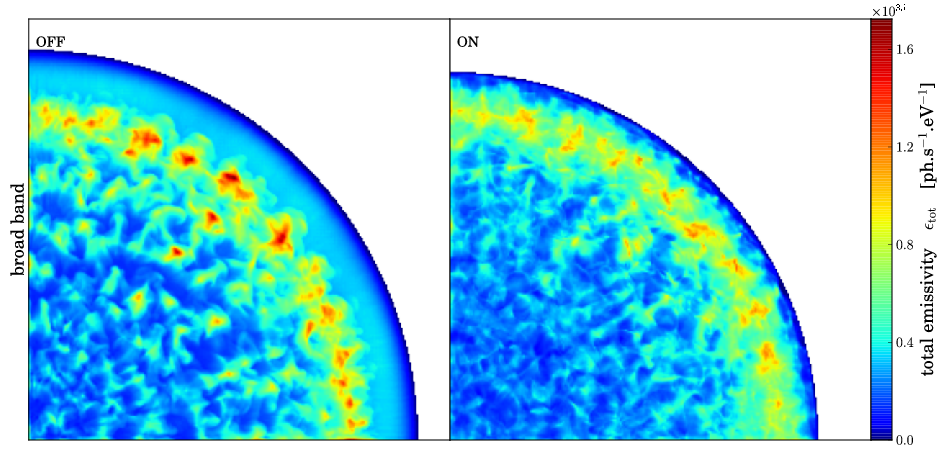}
\caption{Total thermal emissivity. Same as Figure~\ref{fig:prj_THtt_broad}, from a simulation with spatial resolution 4~times lower in each direction.
\label{fig:prj_THtt_broad_lowres}}
\end{figure}

\begin{figure}
\renewcommand{\figwidth}{16cm}
\centering
\includegraphics[width=\figwidth]{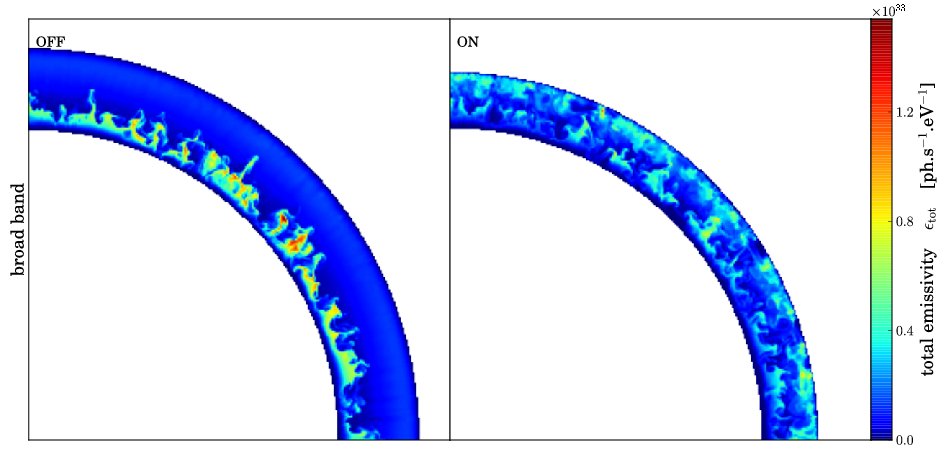}
\includegraphics[width=\figwidth]{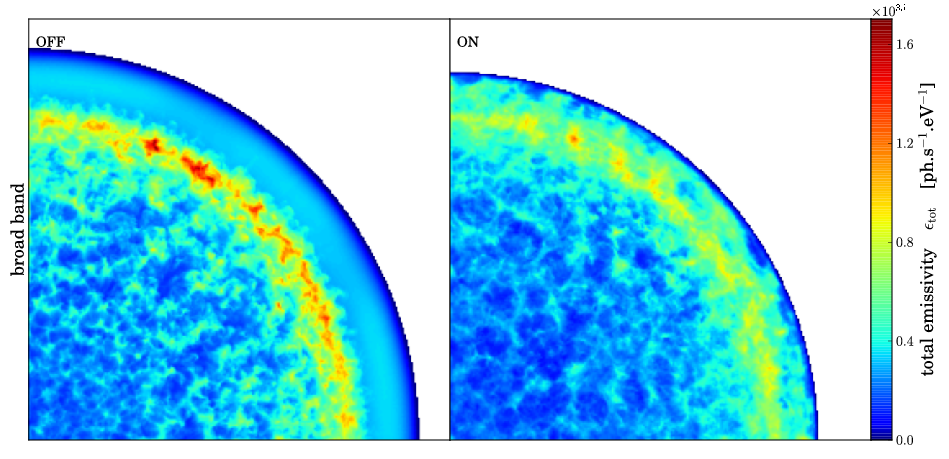}
\caption{Total thermal emissivity. Same as Figure~\ref{fig:prj_THtt_broad}, re-binned by 4x4 blocks. 
\label{fig:prj_THtt_broad_binned}}
\end{figure}

\begin{figure}
\renewcommand{\figwidth}{14cm}
\centering
\includegraphics[width=\figwidth]{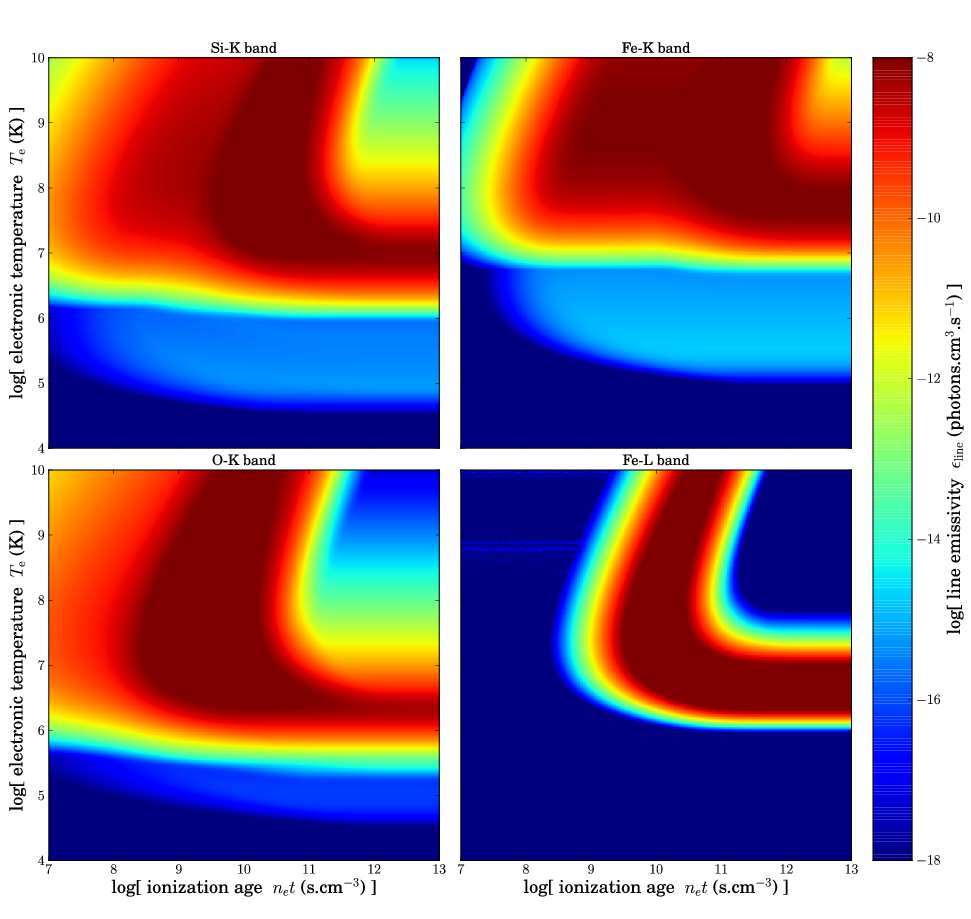}
\caption{Line emissivities in four energy bands:
O-K band: $545-580$~eV, Fe-L band: $730-750$~eV, Si-K band: $1715-1840$~eV, Fe-K~band: $6250-6520$~eV. 
The quantity plotted still has to be multiplied by the density squared to get the actual emissivity per unit volume, hence the units of photons.cm$^3$.s$^{-1}$.
\label{fig:linem}}
\end{figure}

We show here some maps with a coarser spatial grid, to investigate the role of numerical resolution. 
We first note that, regardless of the resolution, we can always make comparisons between the ejecta and the ISM, between cases with and without back-reaction of particles, and see the effects of progressive temperature equilibration and ionization downstream of the shocks. Numerical resolution is a concern mostly because of the fractal-like structure of the Rayleigh-Taylor instabilities: more and more details appear in the fingers as the grid cells become smaller. On Figure~\ref{fig:prj_THtt_broad_lowres} we show the same maps of the broad-band thermal emission as on Figure~\ref{fig:prj_THtt_broad}, with a grid of $256^3$ instead of $1024^3$ (that is $4\times$ times less cell in each of the 3~spatial directions). Although the same overall structure can immediately be recognized, the two sets of maps look different: we can see finer structures at higher spatial resolution. However, the average spectra (not shown here) stay the same within a few \%. 
Here a word of caution on the scales is in order: the maximum flux appears to be higher by more than one order of magnitude on Figure~\ref{fig:prj_THtt_broad_lowres} compared to Figure~\ref{fig:prj_THtt_broad}, simply because of the lower pixel density (divided by $4\times4=16$). If we re-bin the high-resolution maps by $4\times4$ blocks to get the same pixels as on the low-resolution maps, then fluxes agree as can be seen on Figure~\ref{fig:prj_THtt_broad_binned} (fluxes in the re-binned slices are still $4\times$ lower, because the slices are still $4\times$ thinner).
Remarkably, the smoothed maps from the high-resolution simulation contain much more details than the maps from the low-resolution simulation, even though both have the same effective pixel resolution. There are small-scale structures present in the high-resolution simulation, that do not appear in the low-resolution simulation even though there is room for them. This demonstrates that the growth of the Rayleigh-Taylor structures has entered a non-linear regime, where the development of structures at a given scale depends on the existence of structures at other scales. As a consequence, numerical simulations made to match the angular resolution of a particular instrument must actually be substantially over-sampled, in order to catch all the physics possibly observed.


\section{Line emissivities}
\label{sec:app_linem}

We show here for reference some line emissivities, computed from the thermal emission code of Section~\ref{sec:model_SNR_emission}, as a function of temperature and ionization age. The emissivity is integrated over 4~different energy bands, corresponding to lines of Oxygen, Silicon and Iron.
We recall that the electronic temperature $T_{\rm e}$ in the shocked region for the remnant considered is of the order of $10^7$~K: it raises from a few million degrees at the shock fonts to about 20~million degrees in the center. The ionization age goes up to $n_e t \simeq 2.5 \times 10^9\:{\rm cm^{-3}.s}$.

It is apparent that the emissivity depends very strongly on both parameters~$T_{\rm e}$ and~$n_e t$: it varies by ten orders of magnitude over the range considered. We observe the same distinctive curved shape for all elements, but with important differences. For instance, for Iron, triggering the Fe-K line requires high enough temperatures ($T_{\rm e} > 10^7~{\rm K}$), but then is relatively weakly dependent of the ionization age (as long as $n_e t > 10^9~{\rm cm^{-3}.s}$); triggering the Fe-L line requires much stricter conditions (in particular $T_{\rm e}$ must be very close to~$10^7~{\rm K}$ if $n_e t$ is in the range $10^9-10^{10}~{\rm cm^{-3}.s}$). Emission in the O-K and Si-K bands requires less restrictive conditions (it is granted over most of the range of $T_{\rm e}$ and $n_e t$ we are concerned with). These behaviours help the interpretation of the RGB maps and equivalent width maps presented in Section~\ref{sec:results_emission_narrow}.

\newpage

\bibliographystyle{apj}
\bibliography{/Users/gferrand/Documents/references}

\newpage

\end{document}